%% file: main.tex
\documentclass[lettersize,journal]{IEEEtran}
\usepackage{cite}
\usepackage{amsmath,amssymb,amsfonts}
\usepackage{graphicx}
\usepackage{textcomp}
\usepackage{xcolor}
\usepackage{subfigure}
\usepackage{threeparttable}
\usepackage{amssymb,mathrsfs,amsmath}
\usepackage{url}
\usepackage{mdwtab}
\usepackage{booktabs}
\usepackage{array}
\usepackage{mdwmath}
\usepackage{eqparbox}
\usepackage{multirow}
\usepackage{color}
\usepackage{threeparttable}
\usepackage{array}
\usepackage{booktabs}
\usepackage{textcase}
\usepackage{siunitx}
\usepackage{ulem}
\usepackage{hyperref}
\usepackage{ulem}
\usepackage{tikz}
\usepackage{diagbox}
\usepackage[section]{placeins}
\usepackage{algpseudocode}

\algnewcommand{\LeftComment}[1]{\Statex \(\triangleright\) #1}
\usepackage[ruled,vlined,linesnumbered]{algorithm2e}
\let\oldnl\nl
\newcommand{\nonl}{\renewcommand{\nl}{\let\nl\oldnl}}

\def\BibTeX{{\rm B\kern-.05em{\sc i\kern-.025em b}\kern-.08em
    T\kern-.1667em\lower.7ex\hbox{E}\kern-.125emX}}

\hypersetup{
     colorlinks = true,
     linkcolor = blue,
     anchorcolor = red,
     citecolor = magenta,
     filecolor = blue,
     urlcolor = black,
}

\begin{document}

\title{\textsc{TbDd}: A New Trust-based, DRL-driven Framework for Blockchain Sharding in IoT}

\author{
        Zixu Zhang, Guangsheng~Yu, Caijun~Sun*, Xu~Wang, Ying~Wang,
        Ming~Zhang, Wei~Ni, Ren~Ping~Liu,\\
        Nektarios Georgalas, and Andrew Reeves

\thanks{Z. Zhang, X. Wang, and R. Liu are with the Global Big Data Technologies Centre, University of Technology Sydney, Australia, 2007 (e-mail: zixu.zhang@student.uts.edu.au; \{xu.wang-1, renping.liu\}@uts.edu.au)}
\thanks{G. Yu and W. Ni are with the Data61, CSIRO, Sydney, Australia, 2015 (email: \{saber.yu, wei.ni\}@data61.csiro.au)}
\thanks{C. Sun* is the corresponding author with Zhejiang Lab, Hangzhou, China (email: sun.cj@zhejianglab.com)}
\thanks{Y. Wang is with the School of Electronics and Information, Hangzhou Dianzi University, Hangzhou 310018, China (email:90023@hdu.edu.cn)}
\thanks{M. Zhang is with the School of Cyber Engineering, Xidian University, Xi’an, 710071, P.R. China (email: Xidianmingz@gmail.com)}
\thanks{N. Georgalas and A. Reeves are with Applied Research, British Telecom, Martlesham, UK (email: \{nektarios.georgalas, andrew.reeves\}@bt.com; @bt.com)}
}

\markboth{Manuscript Draft}%
{Shell \MakeLowercase{\textit{et al.}}: A Sample Article Using IEEEtran.cls for IEEE Journals}

\maketitle

\begin{abstract}
Integrating sharded blockchain with IoT presents a solution for trust issues and optimized data flow. Sharding boosts blockchain scalability by dividing its nodes into parallel shards, yet it's vulnerable to the $1\%$ attacks where dishonest nodes target a shard to corrupt the entire blockchain. Balancing security with scalability is pivotal for such systems. Deep Reinforcement Learning (DRL) adeptly handles dynamic, complex systems and multi-dimensional optimization.
This paper introduces a Trust-based and DRL-driven (\textsc{TbDd}) framework, crafted to counter shard collusion risks and dynamically adjust node allocation, enhancing throughput while maintaining network security. With a comprehensive trust evaluation mechanism, \textsc{TbDd} discerns node types and performs targeted resharding against potential threats. The model maximizes tolerance for dishonest nodes, optimizes node movement frequency, ensures even node distribution in shards, and balances sharding risks. Rigorous evaluations prove \textsc{TbDd}'s superiority over conventional random-, community-, and trust-based sharding methods in shard risk equilibrium and reducing cross-shard transactions.
 
\end{abstract}

\begin{IEEEkeywords}
blockchain, sharding, collusion attacks, trustworthiness, deep reinforcement learning, internet-of-things
\end{IEEEkeywords}

\IEEEpeerreviewmaketitle

\input{./Sections/1-Introduction}

\input{./Sections/2-RelatedWork}

\input{./Sections/3-SystemModel}
\input{./Sections/4-TBDD}       
\input{./Sections/5-Experiments}
\input{./Sections/6-Conclusions}

\bibliographystyle{./bibliography/IEEEtran}
\bibliography{./bibliography/IEEEabrv}

\vfill

\end{document}

%% file: Sections/1-Introduction.tex
\section{Introduction}
By interconnecting devices, vehicles, and appliances, 
the Internet of Things (IoT) has transformed sectors ranging from smart cities~\cite{qian2019internet}, which optimize traffic and energy use, to autonomous vehicles~\cite{kang2018blockchain} that aim for safer roads, industrial networks~\cite{singh2020deep} that redefine production processes, e-health solutions~\cite{gadekallu2021blockchain} that prioritize patient care, and smart homes~\cite{park2017comprehensive} that enrich daily living. The massive and sensitive nature of the data generated demands strong security and integrity, leading to the integration of IoT and blockchain~\cite{wang2023blockchain, iot-bc-1, iot-bc-2, 8989788, wang2019survey}. This convergence, with its decentralized, immutable, and traceable features, offers a foundational structure for secure data exchanges~\cite{mathur2023survey}. Sharding, a key strategy in blockchain, addresses vast data needs and ensures scalability~\cite{yu2020survey}. By dividing the network into smaller segments, sharding enables simultaneous transaction processing, lightening the load on nodes and increasing transaction throughput.

As the IoT landscape expands, managing and securing vast networks becomes more intricate. However, while various sharding techniques like random-based~\cite{luu2016secure, kokoris2018omniledger, zamani2018rapidchain, wang2019monoxide}, community-based~\cite{zhang2022community}, and trust-based~\cite{yun2019trust} methods enhance the scalability and distribution of dishonest nodes, they remain vulnerable to advanced threats such as adaptive collusion attacks. In such attacks, dishonest nodes manipulate sharding systems by modifying their behavior to focus on specific shards, potentially initiating the $1\%$ attack~\cite{aponte202151}, which significantly compromises the blockchain's security, trustworthiness, and immutability.

Deep Reinforcement Learning (DRL) has become instrumental in the confluence of IoT and blockchain sharding due to its proficiency in handling multiple sharding variables~\cite{yun2020dqn, yang2022sharded,liu2019performance, liu2018blockchain, qiu2019service, qiu2018blockchain}. Amidst the surging blockchain transaction data, DRL provides real-time adaptability, fostering dynamic shard adjustments that enhance resource use and overall system efficiency. However, many DRL-focused studies underestimate the challenges of strategic collusion attacks, where dishonest nodes might disguise their identity or increase cross-shard transactions to hide their intent.

This study investigates the expansion and fortification of blockchain sharding by introducing an innovative framework that utilizes trust tables and DRL. The introduced trust table advances the sharding process by incorporating a multi-dimensional approach to gather feedback within the voting mechanism. This multi-faceted feedback includes direct feedback, indirect feedback, and historical behaviors during the voting process, all contributing to a robust defense mechanism against $1\%$ attack risks. Concurrently, the DRL-driven sharding framework is designed to dynamically allocate shards in real-time, which streamlines node synchronization and minimizes cross-shard transactions (CSTs). This approach also ensures a balanced distribution of dishonest nodes, aiming to decrease the necessity for frequent resharding. The study suggests security thresholds for this DRL approach to enhance training and fortify the sharded blockchain's security.

The primary contributions of this study can be summarized as follows:

\begin{itemize}

\item We propose a secure, scalable, sharded blockchain system architecture for IoT scenarios. Our proposed system effectively mitigates the risks associated with $1\%$ attacks and reduces CST, striking a balance between security and scalability.

\item  We design a novel blockchain sharding system, \textsc{TbDd}, incorporating trust tables to distinguish between dishonest and honest nodes. We consider multiple perspectives for generating a trust table to identify node properties, such as distributed voting, consensus, and node historical behavior analysis.

\item We leverage the DRL framework to monitor the blockchain sharding process continually. By dynamically balancing scalability and security, we achieve high throughput while maintaining the number of dishonest nodes below the security threshold in each shard, ensuring a robust and secure decentralized system.
\end{itemize}

Comparative experiments were conducted simulating strategic collusion attacks across different sharding methods. The outcomes highlight that the \textsc{TbDd} system excels in minimizing CST and ensuring a balanced shard trust distribution. Additionally, the \textsc{TbDd} framework, in the absence of collusion attacks and within tolerable dishonest node levels, offers about a $10\%$ throughput advantage over random-based sharding and a $13\%$ enhancement against the trust-based approach.

The organization of the remaining paper is as follows: Section~\ref{section: related work} reviews relevant literature. Section~\ref{section: system model} introduces the proposed system model and problem definition. Section~\ref{section: TBDD} comprehensively presents the proposed \textsc{TbDd} system. Experiments and assessments are conducted in Section~\ref{section: experiment}. The paper is concluded in Section~\ref{section: conclusion}.

%% file: Sections/2-RelatedWork.tex
\section{Related Work}
\label{section: related work}
In this section, we present the existing related work in the field of blockchain-enabled IoT. Then, we review research on blockchain sharding techniques, including random-based sharding, community-based sharding and trust-based sharding. Additionally, we compare the use of DRL technology in sharded blockchains.

\subsection{Blockchain and IoT}
With the rapid development of blockchain technology, blockchain-enabled IoT has become more secure and reliable. Existing blockchain-enabled IoT systems primarily focus on framework design and consensus protocol design. For instance, Kang \textit{et al.}~\cite{kang2018blockchain} proposed a solution to address security and privacy challenges in Vehicular Edge Computing and Networks (VECONs) based on consortium blockchain and smart contracts. In the medical scenario, Gadekallu \textit{et al.}~\cite{gadekallu2021blockchain} presented a blockchain-based solution to enhance the security of datasets generated from IoT devices in e-health applications. The proposed solution utilizes a blockchain platform and a private cloud to secure and verify the datasets, reducing tampering risks and ensuring reliable results. 
Xu \textit{et al.} proposed a blockchain-enabled system for data provenance with 5G and LoRa network~\cite{9526860,YU2021102492}.

\subsection{Blockchain Sharding}

\subsubsection{Random-based sharding}
In the realm of blockchain sharding research, a series of techniques based on random sharding have been proposed to enhance system scalability and fault tolerance. Elastico~\cite{luu2016secure} is a pioneering protocol that introduced sharding within a permissionless setting, dynamically scaling the network in tandem with the increasing number of participating nodes. OmniLedger ~\cite{kokoris2018omniledger} took this further, ensuring linear scalability through an effective sharding mechanism that seamlessly manages cross-shard transactions. RapidChain ~\cite{zamani2018rapidchain} built upon these concepts with a full sharding solution that streamlines the partitioning of blockchain states and the processing of transactions. Monoxide~\cite{wang2019monoxide} introduced asynchronous consensus areas to boost transaction efficiency, though its static sharding strategy may fall short under dynamically changing network conditions. These innovations have significantly contributed to system robustness by improving randomness and fault tolerance, yet challenges such as frequent node reassignments remain, leading to non-trivial system overheads.

\subsubsection{Community-based sharding}
Community-based sharding is a partitioning method in blockchain networks that divides the system into smaller subsets or shards based on the interactions between nodes. Nodes and transactions are depicted as a graph and are partitioned into smaller subgraphs or shards, each comprising a subset of nodes and transactions. Fynn \textit{et al.}~\cite{fynn2018challenges} investigate the Ethereum blockchain's scalability through sharding implementation. They model the Ethereum blockchain as a graph and analyze different algorithms for graph partitioning, such as the Kernighan-Lin algorithm ~\cite{kernighan1970efficient} and METIS~\cite{karypis1998fast}. Zhang \textit{et al.}~\cite{zhang2022community} proposed community detection-based sharding to enhance sharded blockchain scalability. They used the community detection algorithm to group nodes with frequent trading in the same shard to reduce CSTs. However, this solution demands significant computational resources and high communication costs.

\subsubsection{Trust-based sharding}
Trust-based blockchain sharding divides the network into smaller shards based on node trust. Nodes are evaluated by reputation and behavior, grouping trusted nodes to enhance security and performance and evenly distributing dishonest nodes. Yun \textit{et al.}~\cite{yun2019trust} presented a Trust Based Shard Distribution (TBSD) scheme to enhance system security and prevent collusion, specifically aiming to address the $1\%$ attack. However, their approach does not consider shard load balance and trust difference, potentially leading to system delays. Huang \textit{et al.}~\cite{huang2020repchain} proposed RepChain, a reputation-based blockchain system with sharding for high throughput, security, and node cooperation. It utilizes a double-chain architecture: transaction chain and reputation chain. However, the double-chain architecture chain increases system complexity and resource needs, resulting in additional overhead. Zhang \textit{et al.}~\cite{zhang2023optimized} proposed a new blockchain sharding model to enhance security and efficiency. Their approach considers shard trust, latency, and node count differences, reducing the risk of blockchain failure. However, the effectiveness of their proposed model heavily depends on obtaining accurate information, which can be challenging to obtain in dynamic blockchain sharding scenarios.

\subsection{Deep Reinforcement Learning-based Sharding}
DRL-based sharding solutions have been making strides in the realm of IoT. Liu \textit{et al.}~\cite{liu2019performance} were pioneers, leveraging DRL in a blockchain framework tailored for Industrial IoT (IIoT). While they dynamically adjusted critical parameters, they missed addressing dishonest attacks. Similarly, another work by Liu \textit{et al.}~\cite{liu2018blockchain} harnessed Ethereum and DRL for IIoT data security but sidestepped throughput scalability concerns. On the other hand, Qiu \textit{et al.} presented service-oriented blockchain solutions, using DRL for refined block production and bandwidth allocation~\cite{qiu2019service,qiu2018blockchain}. However, these lacked centralized security measures, making them susceptible to dishonest threats.

Further innovations came from Yun \textit{et al.}~\cite{yun2020dqn}, who introduced a DQN-optimized framework for sharded blockchains. Despite its advancements, the approach overlooked intricate attack strategies in DRL-based sharding and was confined by its central Q-learning reliance. Meanwhile, Yang \textit{et al.}~\cite{yang2022sharded} incorporated K-means clustering in a sharded blockchain, utilizing DRL for optimization. Yet, their methodology was marred by the computational intensity and intricacy of DRL.
Most existing articles~\cite{liu2019performance, liu2018blockchain, qiu2019service, qiu2018blockchain} lack effective analysis of dishonest attacks in IoT using DRL-based blockchain, neglecting the complexities and depth of practical security threats. In contrast, our proposed model analyses the $1\%$ attack problem in the blockchain sharding model and considers strategic collusion attacks involved in existing related studies. By understanding attackers' motivation, we develop proactive defense mechanisms and risk mitigation strategies. In this paper, we design a trustworthy DRL-based blockchain sharding system under IoT scenarios.

%% file: Sections/3-SystemModel.tex
\section{System Model: Mining in Permissioned Shared Blockchain Networks}
\label{section: system model}

In this section, we present the architecture of the proposed sharded blockchain designed to support an IoT network, followed by roles involved in the system, providing a workflow overview and elucidating the system assumptions. 

\subsection{System Overview}
\begin{figure}[t]
	\centering
	\includegraphics[width=1\columnwidth]{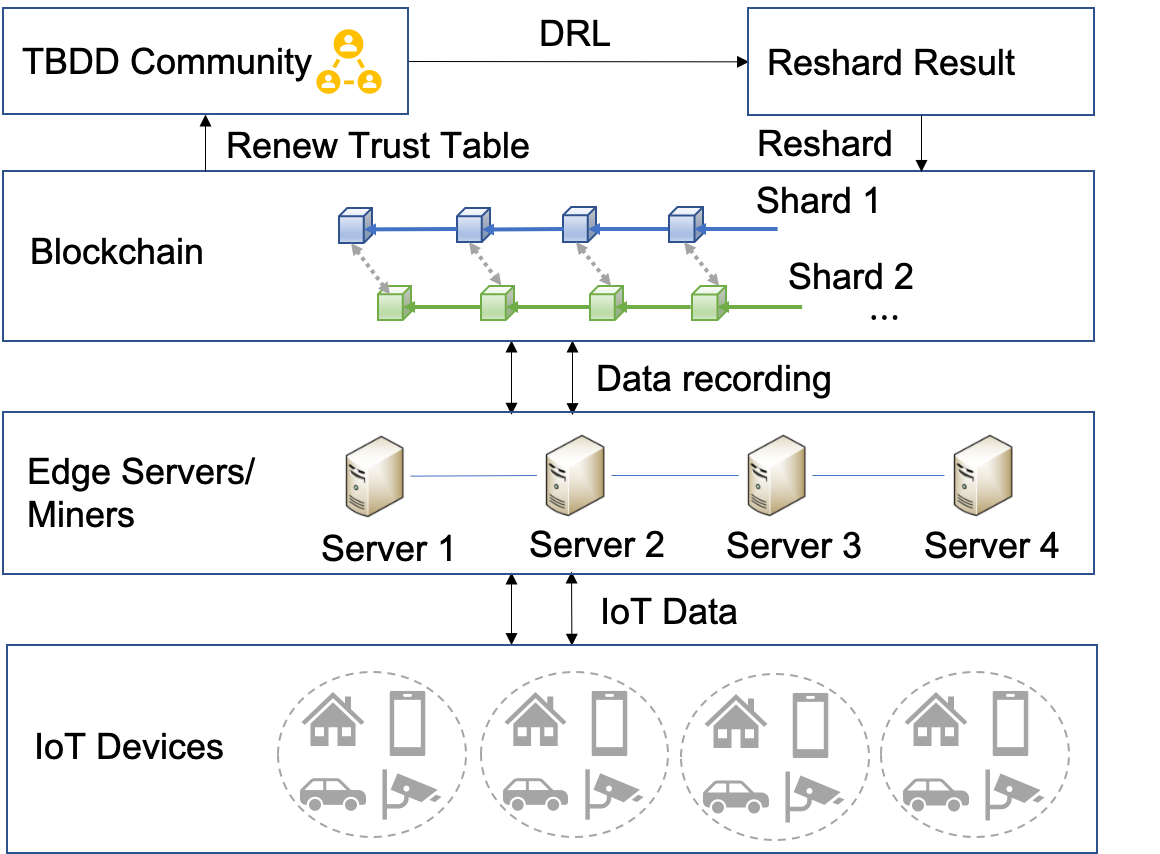}
	\caption{System model.}
        \label{fig.15}
\end{figure}

\begin{figure*}[t]
	\centering
	\includegraphics[width=1\textwidth]{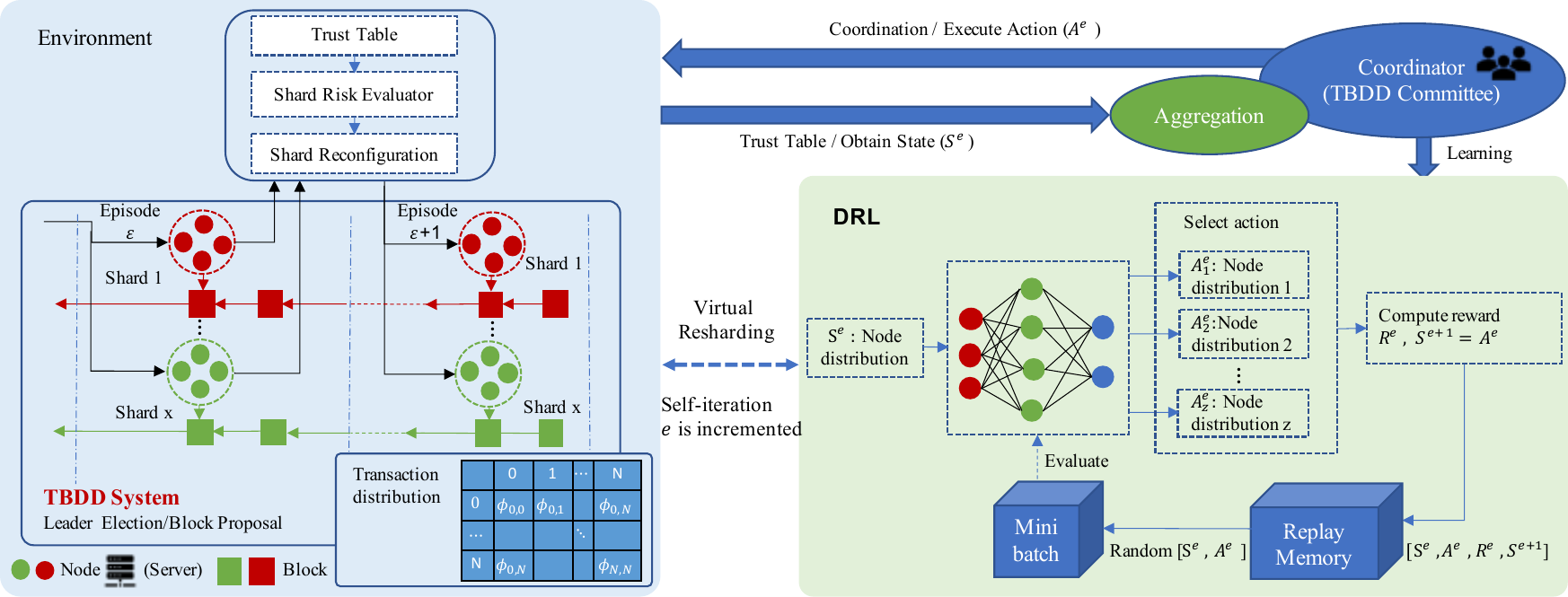}
	\caption{The proposed blockchain sharding system architecture - \textsc{TbDd}.}
        \label{fig.1}
\end{figure*}

\noindent\textbf{System Model.} Fig.~\ref{fig.15} illustrates the proposed sharding framework used in IoT deployments. This framework aims to enhance the scalability and efficiency of blockchain applications within the vast and interconnected realm of IoT devices. Central to the architecture is the sharding mechanism that partitions the broader network into smaller, more manageable shards. Each shard handles a subset of the overall transactions, allowing for simultaneous processing and increasing throughput. Additionally, the design ensures a balanced distribution of nodes across shards to mitigate adaptive collusion attacks.

\smallskip
\noindent\textbf{Architecture.} Fig.~\ref{fig.1} illustrates \textsc{TbDd}'s architecture. Within each shard managed by an edge server, any node can propose a block as a leader. Each node maintains a local trust table with trust scores assigned to other network nodes. We've introduced the \textsc{TbDd Committee} (TC), composed of members democratically selected by the network's users. Drawing inspiration from Elastico\cite{luu2016secure}, the TC ensures decentralized and reliable oversight. This design prevents single points of failure and minimizes risks from a centralized authority.
The TC serves as a decentralized, trusted coordinator, overseeing the user list for the entire network and designating nodes to shards. It aggregates nodes' local trust tables to derive a global trust metric per node, safeguarding trust and node distribution. Essential to \textsc{TbDd} is the resharding mechanism, regularly redistributing nodes among shards. This adaptability ensures the system remains scalable and secure, accommodating a higher transaction volume without compromising on security.

\smallskip
\noindent\textbf{Roles.} In \textsc{TbDd}, the users can take two roles: \textit{validator} and \textit{leader}. A user is a node in the network, which represents a server. All nodes within this network can play the role of validators, participating in the verification process of proposed blocks. There are $N$ participated nodes in the considered blockchains sharding network $\mathbb{N} = \{ 1,\cdots, N \}$. A blockchain sharding network is divided into $D$ shards to verify blocks during the episode of $\varepsilon$. The block proposer is regarded as the leader. The leader broadcasts its proposed block to other peers in the shard for validation. It is important to note that in this system, the role of the leader is trivial because the consensus algorithm is not considered.

\subsection{Workflow Overview}

\begin{figure}[t]
	\centering
	\includegraphics[width=1\columnwidth]{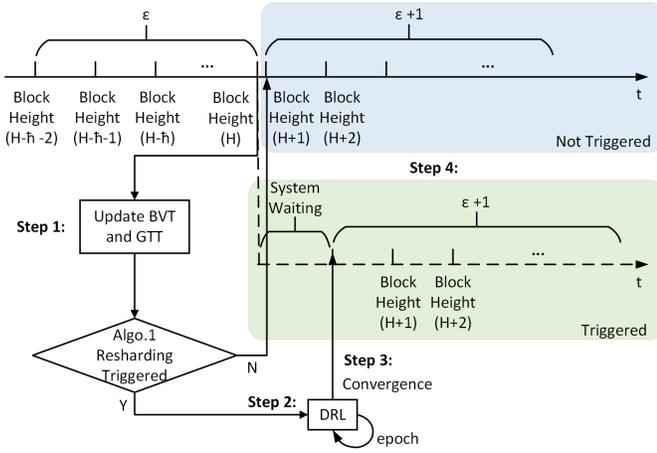}
	\caption{The proposed blockchain sharding system flowchart - \textsc{TbDd}, including four steps. \textcircled{1} Trust table updated: update the Block Verification Table (BVT) and Global Trust Table (GTT), checking whether triggering Algo.~\ref{alg:1}. \textcircled{2} Train the DRL algorithm: use the DRL-based model to virtually resharding through several epochs and output a new node allocation result. \textcircled{3} Update shard allocation: allocate nodes to the shards according to the DRL-based training result. \textcircled{4} Monitor network performance: monitor and step into retraining.}
	\label{fig.2}
\end{figure}

The proposed \textsc{TbDd} framework follows the workflow in Fig.~\ref{fig.2}. 

\noindent\hangindent 1em  \textbf{\textit{Step-1.}}
\noindent\textit{Trust table updated.} 
The first step in the workflow is updating the trust table.  The trust table shows the global trust scores for individual users. Then, the system checks if conditions for triggering Algo.~\ref{alg:1} are met.

\noindent\hangindent 1em  \textbf{\textit{Step-2.}}
\noindent\textit{Train the DRL algorithm.} This step trains the DRL algorithm using the collected network data. This involves running simulations to optimize shard allocation policies based on real-time network conditions. After initiating the DRL training process, the system enters a waiting state and dedicates its computational resources to the training procedure. Note that the TC conducts ``virtual resharding'' trials to evaluate outcomes and rewards for different actions; however, these trials occur solely within the TC, and the final sharding action is not implemented until it has fully converged.

\noindent\hangindent 1em  \textbf{\textit{Step-3.}}
\noindent\textit{Update shard allocation.} TC updates shard allocation policies in real time once the DRL model completes training and allocates nodes to shards based on the result. 

\noindent\hangindent 1em  \textbf{\textit{Step-4.}}
\noindent\textit{Monitor network performance.} As shard allocation policies are updated, monitoring network performance is important. This involves tracking network metrics such as transaction distribution and volume, node location, and colluding risk.

\noindent After \textbf{\textit{Step-4}}, the TC moves on to the retraining phase of the DRL model, incorporating the latest information from the trust table obtained in \textbf{\textit{Step-1}}. Subsequently, the TC updates the shard allocation strategy in \textbf{\textit{Step-3}} based on the new insights gained from the retrained DRL algorithm. This cyclical process follows a similar set of steps, beginning from \textbf{\textit{Step-1}} and progressing through \textbf{\textit{Step-4}}.

\subsection{System Assumptions}
We have several assumptions. These assumptions are considered from multiple perspectives, such as attack and system environment. 

\subsubsection{Attack assumption}
The collusion behavior of nodes refers to the situation where multiple nodes work together to manipulate the network and gain some unfair advantage. This dishonest behavior can overload the network, leading to delays and service disruptions. We design an attack model focusing on collusion attacks in blockchain sharding, and we consider two types of participants: \textbf{dishonest nodes} and \textbf{honest nodes}.

\smallskip
\noindent\textbf{Dishonest nodes.}
Dishonest nodes aim to be allocated in the same shard and then intentionally propose invalid results to honest nodes' blocks. This is achieved by sending many invalid transactions across dishonest nodes, regardless of whether they are cross-shard or intra-shard transactions. Suppose a dishonest node finds no other teammates in the same shard. In that case, it may hide by pretending to be honest and following honest nodes' behavior, avoiding suspicion. Dishonest nodes can also utilize the global trust table's information to enhance their trust by imitating the conduct of honest nodes, selectively endorsing or opposing blocks according to proposers' trust scores.

In \eqref{eq:1}, we normalize the trust score $\mathcal{G}_{i}^\varepsilon$ of the $i$-th node to a range of $\left[0,1\right]$. Let $\mathcal{G}^\varepsilon_{min}$ and $\mathcal{G}^\varepsilon_{max}$ be the minimum and maximum possible raw trust scores. We calculate the current episode's normalized value $\mathbf{G}_i$ by the previous episode's global trust score.

\begin{equation}
    \label{eq:1}
    \mathbf{G}_i = \frac{\mathcal{G}_{i}^{\varepsilon-1} - \mathcal{G}_{min}^{\varepsilon-1}}{\mathcal{G}_{max}^{\varepsilon-1} - \mathcal{G}_{min}^{\varepsilon-1}}.
\end{equation}

We utilize probabilities related to voting behavior, in which the probability of a node voting for a block is denoted as $P_\text{vote}$, while the probability of voting against a block is represented by $P_\text{not\_vote}$. We further distinguish probabilities of voting behavior between dishonest and honest nodes. The probability of dishonest nodes engaging in honest voting and dishonest voting is respectively denoted as $P_\text{vote}^\text{dishonest}$ and $P_\text{not\_vote}^\text{dishonest}$. Let $F$ be the probability of a node's vote failing to reach others due to network issues, with $F \in [0, 1]$. Assume $A$ is the proportion of dishonest nodes in the shard, with $A \in [0, 1]$. We can define a strategy threshold $\tau$, with $\tau \in [0, 1]$, determining when a dishonest node would try to hide by pretending to be honest. If $A < \tau$, the dishonest nodes pretend to be honest and follow honest nodes' behavior; otherwise, they follow the collusion strategy and favor their conspirators. A block verification result, denoted by $U$, takes a $U = 1$ if it matches the local version and $U = 0$ otherwise. We can define a weighting factor $w_G$ based on the normalized trust score $\mathbf{G}$ and a weighting factor $w_U$ based on the block verification result in $U$. We can define the probability distribution for dishonest nodes \eqref{eq:2} and \eqref{eq:3} as follows:

\begin{equation}
    P_\text{vote}^{\text{dishonest}} = 
    \begin{cases}
        (1-F) \cdot w_G \cdot \mathbf{G}_i \cdot  w_U \cdot U, & \text{if}\ A < \tau \\
        (1-F) \cdot \textbf{CollusionStrategy }(\kappa), & \text{otherwise}
    \end{cases}
    \label{eq:2}
\end{equation}

\begin{equation}
    P_\text{not\_vote}^{\text{dishonest}} = 1 - P_\text{vote}^{\text{dishonest}},
    \label{eq:3}
\end{equation}
where $\textbf{CollusionStrategy }(\kappa)$ signifies a strategy in which attackers consistently vote in favor of their teammates while voting against honest leaders with a probability denoted by $\kappa \in [0,1]$. The collusion strategy \eqref{eq:4} is to give valid results to all dishonest partners, attack honest nodes according to a particular proportion, and give them non-valid.

\begin{equation}
     \textbf{CollusionStrategy }(\kappa)= 
    \begin{cases}
        1, & \text{dishonest nodes} \\
        1-\kappa, & \text{honest nodes}
    \end{cases}
    \label{eq:4}
\end{equation}

\smallskip
\noindent\textbf{Honest nodes.}
Honest nodes rely on the global trust table, providing an overview of high-risk or low-risk users without explicitly identifying dishonest nodes. During the intra-consensus phase, honest nodes rely on block verification, voting for a block only if it matches their local version. We propose a composite probability distribution considering the trust-based voting distribution and the block verification distribution. This is achieved by weighting the probability of voting for a block based on the proposer's trust score and adjusting the weight according to the block verification result. This composite distribution allows honest nodes to make more informed decisions when voting for or against proposed blocks, considering both the proposer's trust score and the consistency of the proposed block with their local version. The combined probability distribution for honest nodes \eqref{eq:5} and \eqref{eq:6} can be calculated as follows:
\begin{equation}
    P_\text{vote}^{\text{honest}} = (1-F) \cdot w_G \cdot \mathbf{G}_i \cdot  w_U \cdot U,
    \label{eq:5}
\end{equation}
\begin{equation}
    P_\text{not\_vote}^{\text{honest}} = 1 - P_\text{vote}^{\text{honest}}.
    \label{eq:6}
\end{equation}
The block verification table can be obtained as desired by simulating honest and dishonest nodes' behavior using these attack models. We can analyze the effects of collusion attacks on the overall system's security and performance in blockchain sharding systems.

In our proposed system, along with honest and dishonest nodes, we introduce the concept of high-risk and low-risk nodes. It is important to clarify that high-risk and low-risk nodes differ from honest and dishonest nodes. Our node evaluation principle classifies nodes as high-risk when their trust scores fall below a specific threshold, while nodes with trust scores above this threshold are categorized as low-risk nodes. These classifications do not necessarily mean high-risk or low-risk nodes are inherently honest or dishonest. Instead, they indicate the level of trustworthiness based on our evaluation criteria. By considering the presence of high-risk and low-risk nodes in the network, our system can better identify and respond to potential collusion attacks, improving the security and integrity of the blockchain sharding framework.

\subsubsection{Environment assumption}

The system operates within permissioned blockchain systems, restricting network access to specific nodes. Despite this limitation, the system remains decentralized as it is maintained by the TC, a committee fairly elected by all peers within the network. For the training process to be considered trustworthy, it is important that the TC is reliable and has a transparent record when training the DRL model. Furthermore, Assumptions involve at least a certain number of members per shard, and the claim that the count of dishonest nodes in each shard does not surpass a certain number of the overall node count is based on the Byzantine Fault Tolerance (BFT) principle~\cite{clement2009making}. In case of node failure or a dishonest node attack, the system continues functioning with the remaining nodes.

%% file: Sections/4-TBDD.tex
\section{\textsc{TbDd}: A Trust-Driven and DRL-based Approach to Optimize Throughput and Security}
\label{section: TBDD}

Based on the results and analysis in Section~\ref{section: system model}, we propose \textsc{TbDd}, composed of the Block Verification Table (BVT), Local Trust Table (LTT), and Global Trust Table (GTT), Shard Risk evaluator and Shard reconfiguration.

 \begin{figure}[t]
	\centering
	\includegraphics[width=1\columnwidth]{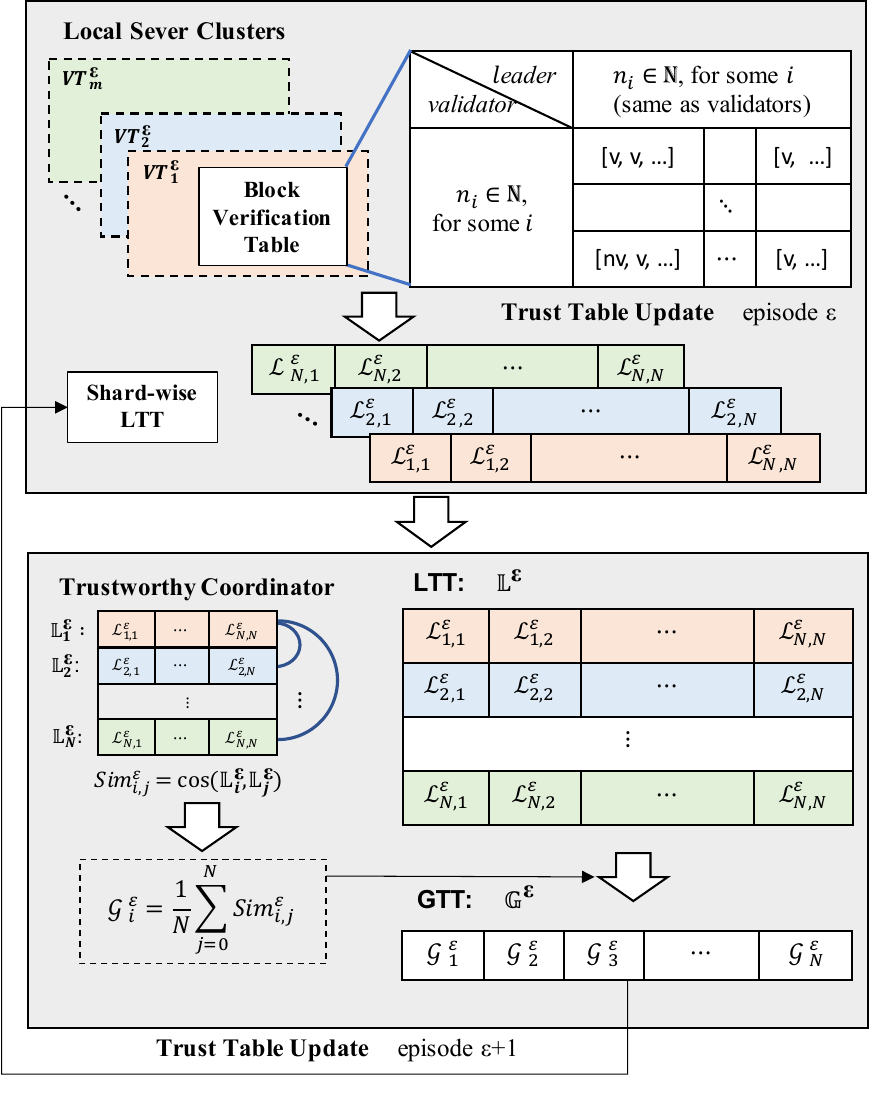}
	\caption{The flow diagram of the proposed computing trust score system comprises BVT, LTT and GTT. }
	\label{fig.3}
\end{figure}

\begin{table}[t]
    \centering
    \caption{Notation and Definition Used in The Trust Table}
    \label{table1}
    \renewcommand\arraystretch{1.2}
    \begin{tabular}{c|p{6cm}}
        \hline
        \textbf{Notation}               & \textbf{Description}  
\\         \hline
$N$     &Total number of nodes in the network
\\
$N_x^\varepsilon$       &Number of nodes in the $x$-th shard in episode $\varepsilon$
\\
$\mathbb{N}$        &The set of all nodes in the network
\\
$n_i \in \mathbb{N}$        &The $i$-th node in the network 
\\
$\iota_j^\varepsilon$       &The times of the $j$-th node
being elected as the leader in episode $\varepsilon$
\\
$D$     & The number of shards 
\\
$\varepsilon$       &The trust table update in episode $\varepsilon$
\\
$e$     &The $e$-th epoch during the DRL iteration
\\

$\mathcal{F}_{i,j}^\varepsilon$     &The indirected feedback of $j$-th leader from the $i$-th node in episode $\varepsilon$
\\

$\hat{\mathcal{F}}_{i,j}^\varepsilon$     &The directed feedback from the $j$-th node to the $i$-th leader in episode $\varepsilon$
\\

$\mathcal{L}_{i,j}^\varepsilon$      &The local trust from the $i$-th node to the $j$-th node in episode $\varepsilon$
\\

$\mathbb{L}_{i}^\varepsilon$     &The local trust table for the $i$-th node in episode $\varepsilon$
\\

$\mathbb{L}^\varepsilon$    &The concatenated local trust tables across all nodes in episode $\varepsilon$
\\

$\mathcal{G}_{i}^\varepsilon$   &The global trust for the $i$-th node in episode $\varepsilon$
\\

$\mathbb{G}^\varepsilon$         &The global trust table in episode $\varepsilon$
\\

$\mathbf{G}_i$         &The normalized trust score of the $i$-th node
\\
$\theta_{x}^\varepsilon$    &The average global trust of $x$-th shard in episode $\varepsilon$
\\
$\bar \theta^\varepsilon$  &The average value of all $\theta_{x}^\varepsilon$
\\
$h_x$       & The list of high-risk nodes in the $x$-th shard
\\
$f_{total}$     &The entire network's fault tolerance threshold for dishonest nodes
\\
$f_{intra}$     &The shard's fault tolerance threshold for dishonest nodes
\\
$\phi_{in}$   & The number of intra-shard transactions (ISTs)
\\ 
$\phi_{cr}$   & The number of cross-shard transactions (CSTs)
\\ \hline 
    \end{tabular}
\end{table}

\subsection{Trust Scheme}
\subsubsection{Block Verification Table (BVT) }
The BVT aims to record the validation results for each shard. The verification table for the $x$-th shard in the $\varepsilon$-th episode is denoted as $VT_x^\varepsilon$. The size of $VT_x^\varepsilon$ is determined by the number of nodes in the shard, with dimensions of $N_x \times N_x$, where $N_x$ represents the number of nodes in the $x$-th shard. The verification result table can be visualized as a two-dimensional matrix where each cell corresponds to the set of verification results generated by the node that proposed a block. The size of each cell in the matrix corresponds to the number of times the leader produced blocks. During the trust table update episode, assume that there are sufficient votes to guarantee that each node within the shard will be elected as a leader at least once, resulting in block production. In this assumption, the leader can expect to obtain a minimum of one ballot from himself. Furthermore, we assume that the consensus process complies with the weak synchrony assumption~\cite{castro1999practical}, indicating a finite upper limit on message delays.

\subsubsection{Trust Table}
In \textsc{TbDd}, two distinct trust tables are utilized: LTT and GTT. These tables are dynamic and regularly refreshed to capture the latest data on each node’s performance and behaviors within the network. By constantly monitoring and evaluating the LTT and GTT, the \textsc{TbDd} system can make strategic and informed shard allocation decisions, ensuring a reliable and secure network operation.

The local trust table is denoted as $\mathbb{L}^\varepsilon$. Each row of this table is specifically assigned to a node within the shard, and these nodes are represented as $1\times N$ entries within the table, which we term as the local trust table of the $i$-th node at the $\varepsilon$ epoch, denoted as $\mathbb{L}_{i}^\varepsilon$. When these individual local trust tables are concatenated, we represent it as $\mathbb{L}^\varepsilon$, forming a comprehensive $N \times N$ table where $i,j \leq N$. Within this table, each element represents the trust score sent from the $i$-th node to the $j$-th node, denoted as $\mathcal{L}_{i,j}^\varepsilon$. The computation of each element in the LTT of $i$-th node is shown in \eqref{eq:7}:
\begin{equation} 
\label{eq:7}
   \mathcal{L}_{i,j}^\varepsilon = \begin{cases}
    \alpha \mathcal{F}_{i,j}^\varepsilon + \beta \hat{\mathcal{F}}_{i,j}^\varepsilon + \mu 
    \mathcal{G}_{i}^{\varepsilon-1}, \\ \qquad\qquad\qquad\quad \text{if } n_i,n_j \text{ in the same shard}\\
    \mathcal{G}_{i}^{\varepsilon-1}, \\ \qquad\qquad\qquad\quad \text{if } n_i,n_j \text{ in different shards}
   \end{cases}
\end{equation}
\noindent when calculating the trust score in the same shard, three feedbacks are included: Indirected feedback $\mathcal{F}_{i,j}^\varepsilon$, Directed feedback $\hat{\mathcal{F}}_{i,j}^\varepsilon$ and Global trust score from the last episode $\mathcal{G}_{i}^{\varepsilon-1}$. Each component has a distinct proportion represented by $\alpha, \beta, \mu$. These proportions sum up to $1$ (i.e., $\alpha + \beta + \mu = 1$). Only the global trust score from the previous episode is considered for calculating the trust scores of nodes in different shards.

\smallskip
\noindent\textbf{Indirected feedback of each leader.}  
The proportion of verification that the $j$-th node passes when he is the leader at $\varepsilon$ episode is shown in \eqref{eq:8}:
\begin{equation} 
\label{eq:8}
V_{j}^\varepsilon = \frac{\sum\nolimits_{{i}=1}^{N_x^\varepsilon}v_{{i},{j}}^\varepsilon }{\iota_{j}^\varepsilon N_x^\varepsilon},
\end{equation}
where $\iota_{j}^\varepsilon$ represents the times of the $j$-th node being elected as the leader in the $\varepsilon$-th episode, and $v_{{i},{j}}^\varepsilon$ signifies the total number of valid votes $v$ cast by the $i$-th node for the $j$-th leader. Then, the indirected feedback $\mathcal{F}_{i,j}^\varepsilon$ is calculated as follows \eqref{eq:9}:
\begin{equation} 
\mathcal{F}_{i,j}^\varepsilon = 
\gamma V_{{j}}^\varepsilon +  \frac{\gamma^2}{N_x^{\varepsilon}-2} \sum\limits_{p \in {\{{i},{j}\}^\text{C}}} \delta_{{p},{j}}^\varepsilon \left( V_{p}^{\varepsilon} \right)^{\iota^\varepsilon_{{j}}-\delta_{{p},{j}}^\varepsilon + 1},
\label{eq:9}
\end{equation}
where the node, denoted as $n_p$, provides indirect feedback and participates in the voting process for the current leader $l_j$. The $n_p$ node does not include block proposer $l_j$ and the voter $n_i$ itself. ${\delta^\varepsilon_{p,j}}$ refers to the ratio of non-empty votes (both valid and non-valid votes) cast from the $p$-th node for the $j$-th leader. $\gamma$ is a discount rate similar to that used in reinforcement learning.

\smallskip
\noindent\textbf{Directed feedback of each leader.} The direct feedback of trust score is calculated as shown in \eqref{eq:10}:

\begin{equation} 
\label{eq:10}
\hat{\mathcal{F}}_{i,j}^\varepsilon = \frac{v_{{j},{i}}^\varepsilon} {\iota_{i}^\varepsilon},
\end{equation}
where $v_{{j},{i}}^\varepsilon$ denotes the count of valid votes $v$ cast by the $j$-th node for the $i$-th node when $n_i$ is leader. $\iota_{i}^\varepsilon$ indicates that the number of times of the $i$-th node being elected as the leader during the $\varepsilon$-th episode.

\smallskip
\noindent\textbf{Global trust of each leader from the history.} The historical trust of each leader 
$\mathcal{G}_{i}^{\varepsilon-1}$ is inherited from the last episode.

The global trust table is denoted as $\mathbb{G}^\varepsilon$. Inspired by federated learning principles, the coordinator enhances credibility by updating the LTT according to the GTT's outcome after every iteration. 

\smallskip
\noindent\textbf{Cosine similarity calculation.} Comparisons of cosine similarity among LTT rows reflect deviations in node-scoring behavior \eqref{eq:11}:

\begin{equation}
    \label{eq:11}
    Sim^\varepsilon_{i,j}=cos(\mathbb{L}^\varepsilon_{i},\mathbb{L}^\varepsilon_{j}), \quad i,j<N.
\end{equation}

The global trust score for each node is determined by computing the mean of the cosine similarity between the node's scoring behavior in the LTT and the entire LTT \eqref{eq:12}:
\begin{equation}
    \label{eq:12}
    \mathcal{G}_{i}^\varepsilon = \frac{1}{N}\sum_{j=1}^{N} Sim^\varepsilon_{i,j},
\end{equation}
$\mathcal{G}^\varepsilon$ is a $1 \times N$ vector capturing the global trust, where element $\mathcal{G}_{i}^\varepsilon$ is the global trust of the $i$-th node in the current shard.

\subsection{Resharding Trigger: The Shard Risk Evaluator}
The resharding process is triggered when certain conditions are met, leading to the trigger phase (the green block in Fig.~\ref{fig.2}), as shown in Algo.~\ref{alg:1}. We define a global trust threshold $\rho_{t}$ to differentiate between low-risk and high-risk nodes. Nodes are high-risk and possibly dishonest if their $\mathcal{G}^\varepsilon$ are lower than $\rho_{t}$. Nodes are \textit{more likely} to be honest if their global trust exceeds $\rho_{t}$. 
We define the fault tolerance threshold $f_{intra}$ within each shard.
If the number of dishonest nodes exceeds the threshold $f_{intra}$ in the shard, the shard is labeled as corrupted and triggers resharding. Furthermore, resharding is also triggered when the ratio of CST surpasses the setting threshold $\rho_{cr}$. The ratio of the CST is calculated as follows:
\begin{equation}
    \label{eq:13}
    \varphi_{cr} = \frac{\phi_{cr}} {\phi_{cr} + \phi_{in}},
\end{equation}
where $\phi_{cr}$ represents the CST count, as given by:
\begin{equation}
    \label{eq:14}
    \phi_{cr} = \frac{1}{2} \left (\sum_{i=1}^{N} \sum_{j=1}^{N} \phi_{i, j} - \sum_{x=1}^{D} \sum_{i=1}^{N_x^\varepsilon} \sum_{k=1}^{N_x^\varepsilon} \phi_{i, k} \right ),
\end{equation}
where $\phi_{i, j}$ represents the transaction count between the $i$-th and $j$-th nodes in the network. $\phi_{i,k}$ represents the transaction count between the $i$-th node and the $k$-th node in the shard, which equals the sum of the Intra-Shard Transaction (IST). The $\phi_{cr}$ is obtained by subtracting the IST count from the total transactions count among network nodes.

\begin{algorithm}[t]  
\footnotesize
\caption{Shard risk evaluator}
\label{alg:1}
\LinesNumbered

\KwIn{\\
\quad $x$: The $x$-th shard;\\
\quad $N_x^\varepsilon$: Number of nodes in the $x$-th shard at $\varepsilon$ episode;\\
\quad $\mathcal{G}_{i}^\varepsilon$: Global trust of the $i$-th node;\\
\quad $f_{intra}$: The faulty tolerance of dishonest nodes in any shard;\\
\quad $\varphi_{cr}$: The CST ratio since last episode;\\
\quad $\rho_{t}$: The threshold of differentiating a low-risky or high-risky node in terms of trust score;\\
\quad $\rho_{cr}$ : The threshold in terms of CST;\\
}

\KwOut{\\
\quad \{\textit{Trigger}, \textit{Not trigger}\}\
}


$\rho_{cr} := 0$ 

$h_x := []$ \textbf{for each shard} $x$

\For{$x \in [1, D) $ }
    {
    \For{$n_i\in x \textbf{ and } i \in [1, N) $} 
    {   
        \If{$\mathcal{G}_{i}^\varepsilon <  \rho_{t} $}
        {$h_x\leftarrow n_i$}
    }
    \If {$ \left| h_x \right| / N_x > f_{intra}$}
    {\textbf{return} \textit{Trigger}}     
}

\If {$ \varphi_{cr} >\rho_{cr}$}
            {\textbf{return} \textit{Trigger}}

\Return \textit{Not Trigger}

\end{algorithm}

\subsection{DRL Framework}
\subsubsection{Optimizations, Rewards, and DRL}
The DRL-based model enhances blockchain sharding systems by dynamically allocating nodes depending on network conditions and automating complex decision-making procedures. The reward function optimizes node allocation while maintaining security constraints through DRL adaptation. In \textsc{TbDd}, the agent aims to maximize earnings by calculating the objective function that is composed of $6$ reward components, where $E_{a}, E_{b}, \lambda_a, \lambda_a$, and $\lambda_c$ are all constant.

\smallskip
\noindent\textbf{Shard load balance ($\xi$).}
As shown in~(\ref{eq:15}), we aim to distribute the number of nodes across each shard evenly. If there is a significant disparity in the node count per shard, resharding is augmented to counteract potential $1\%$ attacks~\cite{han2022analysing}.
\begin{equation}
    \label{eq:15}
    \xi=\left\{\begin{array}{ll}
    E_{a}, & \text {shard load balance}  \\
    -E_{a}, & \text {shard load unbalance}
\end{array}\right.
\end{equation}

\smallskip
\noindent\textbf{Corrupted shards portion ($\varrho$).}
As shown in ~(\ref{eq:16}), the agent receives a reward if no shard is occupied. Otherwise, the agent receives a penalty.
\begin{equation}
    \label{eq:16}
    \varrho=\left\{\begin{array}{ll}
    E_{b}, & \text {no shard is corrupted} \\
    -E_{b}, & \text {at least one shard is corrupted}
\end{array}\right.
\end{equation}

\smallskip
\noindent\textbf{CST ratio ($\eta$).}
As shown in ~(\ref{eq:17}), if the CST ratio is less than the specified threshold $\rho_{cr}$, the agent gets rewarded; otherwise, it receives a penalty.

\begin{equation}
    \label{eq:17}
    \eta= \lambda_a \left (\rho_{cr}-\varphi_{cr} \right ) \lambda_b^{\lambda_c \left | \varphi_{cr}- \rho_{cr} \right | },
\end{equation}

\smallskip
\noindent\textbf{Nodes shifting ratio ($\psi$).}
As shown in ~(\ref{eq:18}), if a node switches the shard, its action is noted as 1; otherwise, it is 0. The overall count of shifted nodes corresponds with the penalty score. The more nodes relocate, the more computational resources are consumed by shard synchronization, resulting in a higher level of punishment.
\begin{equation}
    \label{eq:18}
    \psi = \frac{1}{N} \sum_{j=1}^{N} \upsilon_{j}, \quad
    \upsilon_{j} =\left\{\begin{array}{ll}
    1, & \text { nodes moving } \\
    0, & \text { nodes staying }
    \end{array}\right.
\end{equation}

\noindent\textbf{Intra-shard's trust variance ($\Omega_{in}$).}
As shown in~(\ref{eq:19}), trust variance represents the trust distribution within each shard. A larger trust variance of a shard indicates a better distinction between trustworthy and untrustworthy participants. Thus, a larger intra-shard trust variance collected by the agent serves as a reward to indicate a clearer differentiation between honest and dishonest nodes.

\begin{equation}
    \label{eq:19}
    \Omega_{in} = \frac{1}{D} \sum_{x=1}^{D} \frac{1}{N_{x}^\varepsilon} \sum_{k=1}^{N_{x}^{\varepsilon}} \left | \mathcal{G}_{k}^\varepsilon - \theta_{x}^{\varepsilon} \right |^2,
\end{equation} 
where $\mathcal{G}_{k}^\varepsilon$ is the global trust value of nodes in the $x$-th shard and $\theta_{x}^{\varepsilon}$ is the average of global trust in the $x$-th shard. 

\smallskip
\noindent\textbf{Cross-shard's trust variance ($\Omega_{cr}$).}
As shown in~(\ref{eq:20}), it represents the deviation of trust value among different shards. A minor cross-shard trust variance indicates a more uniform distribution of dishonest nodes.

\begin{equation}
    \label{eq:20}
    \Omega_{cr} = \frac{1}{D} \sum_{x=1}^D \left | \theta_{x}^{\varepsilon} -  \bar\theta^{\varepsilon}  \right |^2,
\end{equation} 
where $\bar\theta^{\varepsilon}$ is the average value of all $\theta_{x}^{\varepsilon}, x \in D$. 
Thus, the objective function can be defined as:
\begin{equation}
\label{eq:21}
    R = \xi +\varrho + \eta - \psi + \Omega_{in} -\Omega_{cr},
\end{equation}

\begin{equation}
    \begin{aligned}
    \label{eq:22}
    &\text{Let } \mathbf{R} = [\xi, \varrho, \eta, \psi, \Omega_{in},\Omega_{cr}],\\
    &\text{objective: } \max_{\mathbf{R}} \sum^{e_{max}} R(\mathbf{R}),\\
    &\text{s.t. } \quad |h_x| < N_x f_{intra},\\ 
    &\quad\quad\quad N_x \geq N_{min},
    \end{aligned}
\end{equation}
where $\xi$ is the balance reward for shard load; $\varrho$ is the reward for the number of corrupted shards; $\eta$ signifies the reward for low-level CST; $\psi$ indicates the reward for the number of shifted nodes; $\Omega_{in}$ stands for the reward of intra-shard trust variance; and $\Omega_{cr}$ represents the reward of cross-shard trust variance. We define $\mathbf{R}$ as the set of all rewards. Restrictions of the objective function $R$ ensure the dishonest node count stays below the shard's fault tolerance. $N_{min}$ denotes the minimum node requirement for each shard. Each shard mandates a minimum of four nodes in our setting, i.e., $N_{min} = 4$.

\subsubsection{DRL-based Sharding Optimization Model}

We use the DRL model to assist in the blockchain reconfiguration process. The agent of the DRL model is acted by the TC, a committee elected by all peers in the network. The agent obtains the state from the node allocation. Then, the agent gains the reward by virtually resharding, deciding the optimal allocation strategy and executing the action for the new node allocation.

\smallskip
\noindent\textbf{Agent:}
The agent is perceived as the TC, which consists of several nodes. These nodes implement the PBFT protocol to achieve consensus, representing the collective action of the TC. The agent executes a learning process and decides shard allocation based on real-time network conditions.

\smallskip
\noindent\textbf{Environment:}
As illustrated in Fig.~\ref{fig.1}, the environment is viewed as a black box that executes the \textbf{Action} of the agents and obtains the \textbf{State}. In this paper, the sharding reconfiguration process in the blockchain sharding network is the environment.

\smallskip
We consider that the agent obtains a state $s \in \mathcal{S}$ based on the node allocation at the current episode $\varepsilon$. $s$ denotes the distribution of all nodes across various shards in the current episode, while $m$ precisely identifies the specific node present within a particular shard.
\begin{equation}
    \label{eq:23}
    s = \left [ m_{x,1}, \cdots, m_{x,i}, \cdots, m_{x,N} \right ], x \in D, i \in N,
\end{equation}
where $m_{x,i}$ is the $x$-shard to which the $i$-th node belongs.



\smallskip
\noindent\textbf{Actions ($\mathcal{A}$).} 
The agent executes an action $a \in \mathcal{A}$ based on its decision produced by its local DRL-based learning during the current episode $\varepsilon$. The valid action space for the agent is formatted identically to $\mathcal{S}$, as given by
$\mathcal{A}=\mathcal{S}$ with $s^{\varepsilon+1}=a^{\varepsilon}$.

\smallskip
\noindent\textbf{Policy ($\pi$).}
A policy determines what action the agent should take next based on a set of rules. In our case, the process of nodes assigned to different shards is based on the policy that is continually updated and trained.  i.e., $ \pi (s, a): \mathcal{S} \rightarrow \mathcal{A}$.

\smallskip
\noindent\textbf{Reward function ($r$).}
The reward function is inherited from the objective function, $r: R(\xi, \varrho, \phi_{cr}, \psi, \Omega_{in}, \Omega_{cr} )$.

DRL is a versatile method that allows agents to make decisions in dynamic environments effectively. Its ability to handle complex data and its proactive defense against malicious attacks make it an ideal solution for managing node and transaction interactions in sharding systems. The sharding optimization problem, which involves assigning each node to a specific shard, is a known non-convex, NP-hard problem due to the binary nature of decision variables~\cite{9500403}. This makes DRL a more suitable approach to address this challenge than traditional methods such as convex optimization, underlining its significance in managing the complexities of sharded blockchain systems.

%% file: Sections/5-Experiments.tex
\section{Experiment and Evaluation}
\label{section: experiment}

We experimentally assess the DRL-based sharding approach regarding convergence performance and stability under various environment settings, including the number of shards, the number of nodes, the resource distribution, and other practical settings. Note that the parameters in the following experiments align with real-world IoT scenarios, such as Mobile Edge Computing (MEC), where edge servers on vehicles or drones collect data from end devices such as sensors. These servers initiate re-sharding to optimize data processing by redistributing workload when moving across regions. 


\begin{table}[t]
    \centering
    \caption{Hyperparameters}
    \label{table2}
    \renewcommand\arraystretch{1.5}
    \resizebox{\columnwidth}{!}{%
    \begin{tabular}{c|l|c}
        \hline
        \textbf{Notation} & \textbf{Description} & \textbf{Value} \\ \hline
        $e_{max}$       & The number of epoches in DRL     & $\left [ 30, 100\right ]$ \\ 
        $F$        & The probability of node failing to vote      & $20\%$ \\ 
        $A$         & The fraction of dishonest nodes within one shard & $\left [ 0, 30\right ]$ \\ 
        $\tau$         & The threshold for triggering colluding strategy    & $10\%$ \\
        $\gamma$        &The discount factor for calculating $OT$ of each leader        & $0.9$ \\
        $\rho_{t}$          &The threshold of differentiating low-risky nodes and & 0.67 \\
        & high-risky nodes in terms of trust& \\
        $f_{intra}$      & The threshold of faulty tolerance within one shard & $\lfloor (N_x-1)/3 \rfloor$ \\ 
        $\rho_{cr}$       & The threshold of triggering resharding in terms of CST ratio & 0.4 \\ 
        \hline
        
    \end{tabular}%
    }
\end{table}

\subsection{Experiment Framework}

An experimental framework is implemented in a local environment with 2.00 Ghz, 2 $\times$ 40 cores, 2 $\times$ Intel(R) Xeon(R) Gold 6138 CPU, NVIDIA PCIe A100, 8 $\times$ 40GB GPU, 250GB memory and 1000Mb/s to evaluate a proposed blockchain sharding scheme. We create a virtual machine using Ubuntu 18.04.6 LTS in Python 3.7.13 and Pytorch 1.13.1. In this setting, we use the discrete DRL algorithms, DQN and PPO, and run over $30-100$ epochs. We set the range of total nodes from $0-16$ in our environment. The transaction distribution model between nodes follows the normal distribution. The trustworthy coordinator is implemented by deploying the smart contract. 

In our experimental setup, we assess the blockchain sharding system \textsc{TBDD} against other sharding techniques such as random-based sharding~\cite{luu2016secure}, community-based sharding~\cite{zhang2022community}, and trust-based sharding~\cite{yun2019trust}. We account for dishonest nodes in the range of $0-5$, which employ collusion strategies during block verification. These nodes may produce deceptive block verifications and send CSTs across different shards. Initially, these dishonest nodes are scattered randomly across shards. To model interactions between dishonest nodes in distinct shards, positive noises are introduced to their transaction counts. The resulting transaction distribution table is influenced by normally distributed transactions. The experiments' hyperparameters are detailed in Table~\ref{table2}.

As the intra-shard fault tolerance threshold $f_{intra}$ is set to $\lfloor (N_x-1)/3\rfloor$, we also evaluate the relationship between the total number of nodes $N$ in the network, the number of shards $D$, and the total fault tolerance $f_{total}$, as given by
\begin{equation}
    \label{eq:24}
    f_{total} = \left \lfloor \left( \left\lfloor \frac{N}{D} \right\rfloor - 1 \right) / 3 \right\rfloor \times D,
\end{equation}
by which one can realize whether the entire network has failed.

\subsection{Experiment Results}
In the experimental evaluations, we compared the performance of the \textsc{TBDD} system with random-based, community-based, and trust-based sharding approaches using metrics such as CST ratio, shard risk variance, corrupted shard number, and convergence speed. Through these evaluations, we validate the effectiveness and robustness of our proposed approach. In the following figures, we label each figure in terms of the number of nodes $N$, shards $D$, and dishonest nodes $h$, respectively.

\begin{figure*}[t]
    \centering
    \setlength{\belowcaptionskip}{-0.3cm} 
    \subfigure[$N=16, D=2, h=4$.]{\label{fig.5a}\includegraphics[width=2.3in]{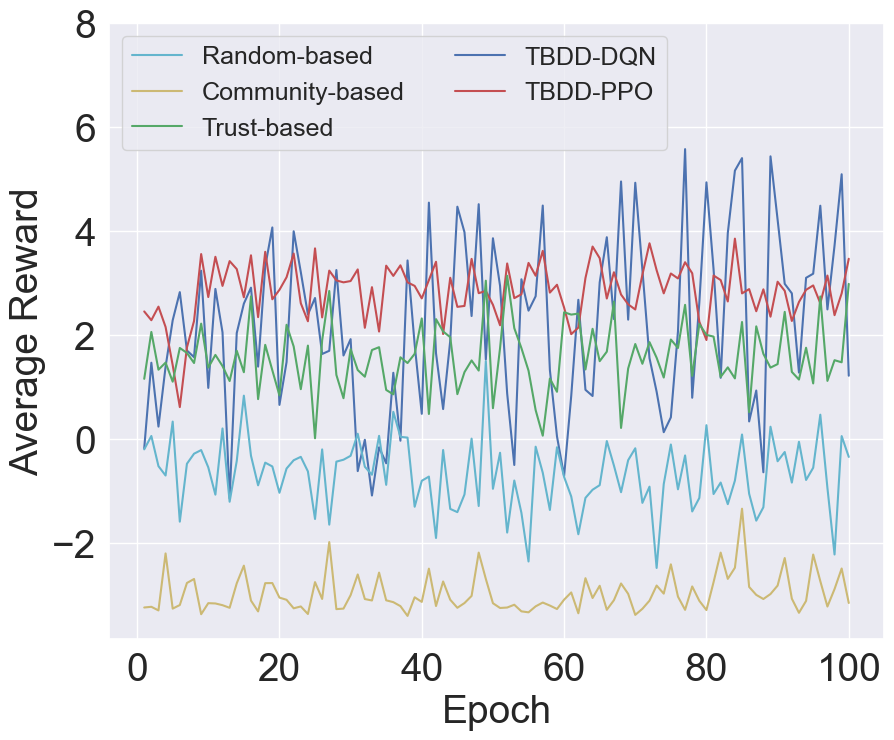}}
    \subfigure[$N=\lbrace10,12,14,16\rbrace, D=2, h=2$.]
    {\label{fig.5b}\includegraphics[width=2.3in]{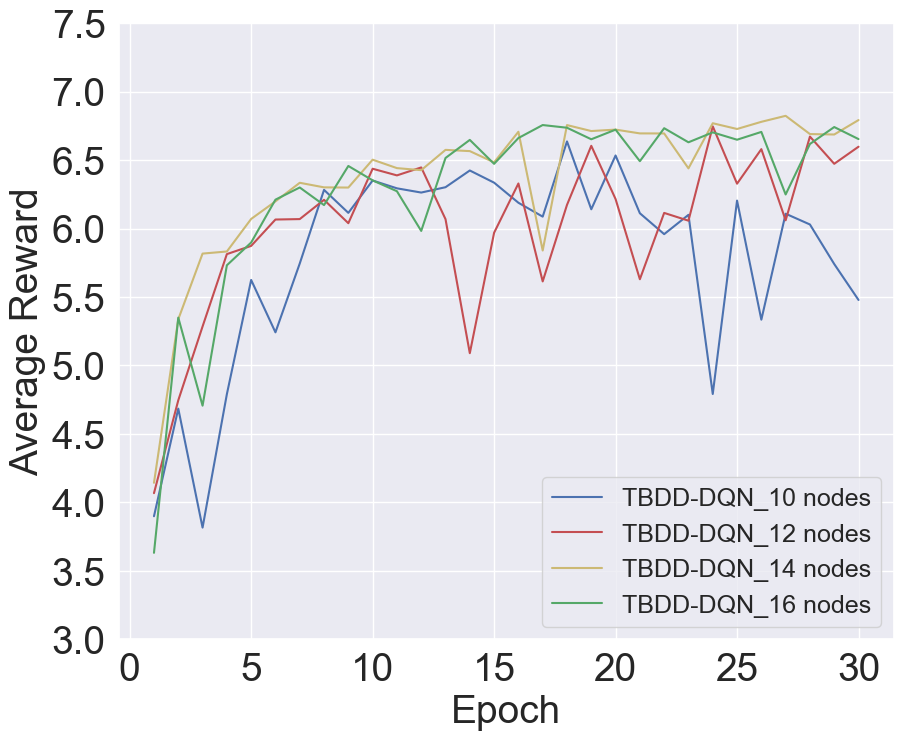}}
    \subfigure[$N=\lbrace10,12,14,16\rbrace, D=2, h=2$.]
    {\label{fig.5c}\includegraphics[width=2.3in]{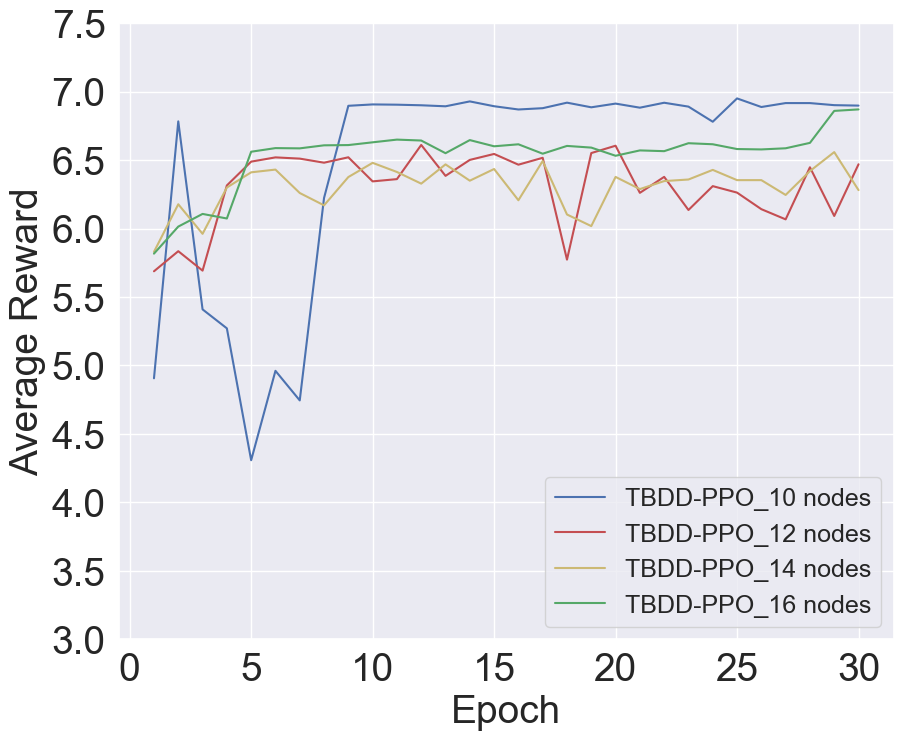}}
    
    \caption{Fig.~\ref{fig.5a} represents the comparison among Random-based, Community-based, Trust-based, \textsc{TBDD-DQN} and \textsc{TBDD-PPO} across $100$ epochs. Figs.~\ref{fig.5b} and \ref{fig.5c}  represent the reward performance between \textsc{TBDD-DQN} and \textsc{TBDD-PPO} across varying node numbers in the environment settings across $30$ epochs, respectively.}
    \label{fig.5}
\end{figure*}

\begin{figure*}[t]
    \centering
    \vspace{-0.3cm}
    \setlength{\belowcaptionskip}{-0.3cm} 
    \subfigure[Random-based]{\label{fig.6a}\includegraphics[width=0.19\linewidth]{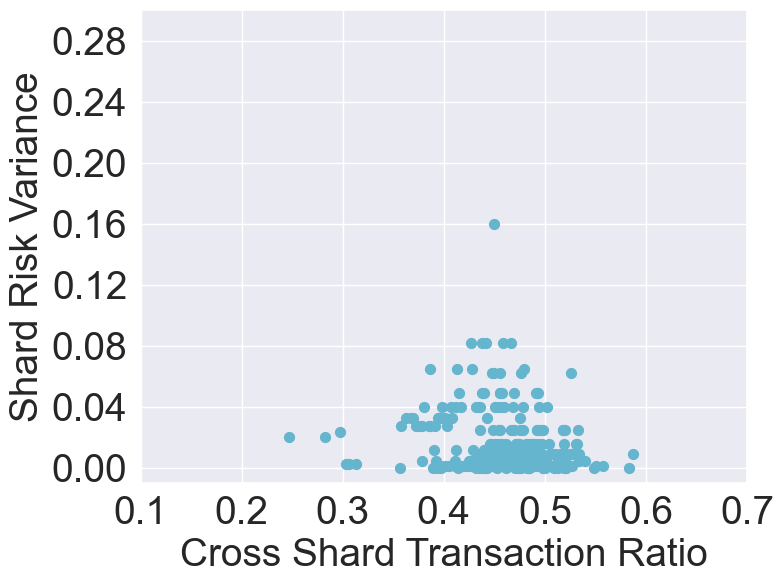}}
    \subfigure[Community-based]{\label{fig.6b}\includegraphics[width=0.19\linewidth]{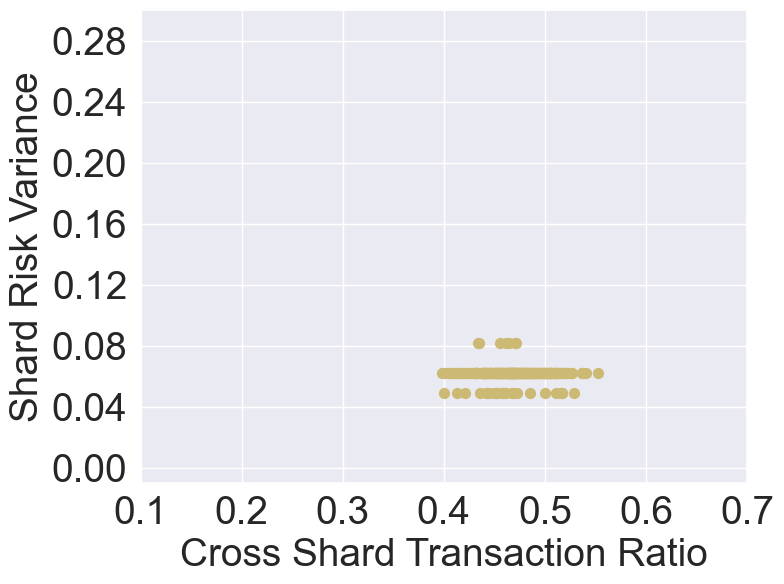}}
    \subfigure[Trust-based]{\label{fig.6c}\includegraphics[width=0.19\linewidth]{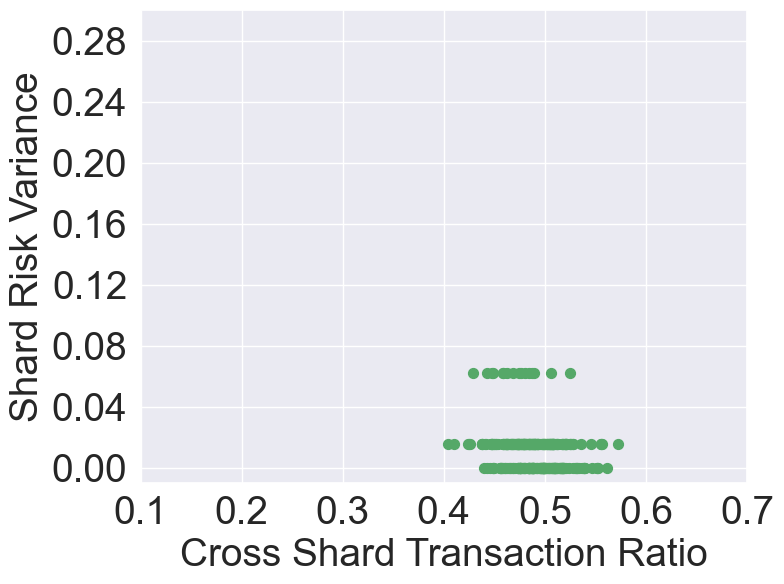}}
    \subfigure[TBDD-DQN]{\label{fig.6d}\includegraphics[width=0.19\linewidth]
    {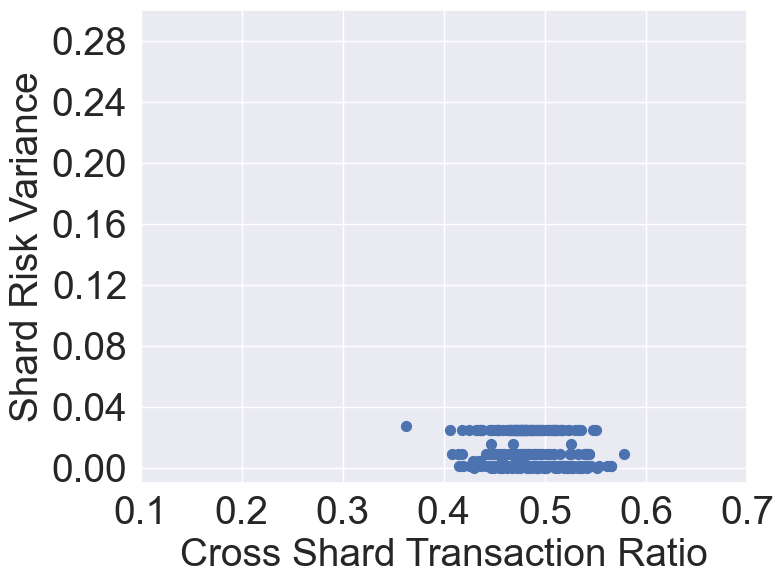}}
    \subfigure[TBDD-PPO]{\label{fig.6e}\includegraphics[width=0.19\linewidth]
    {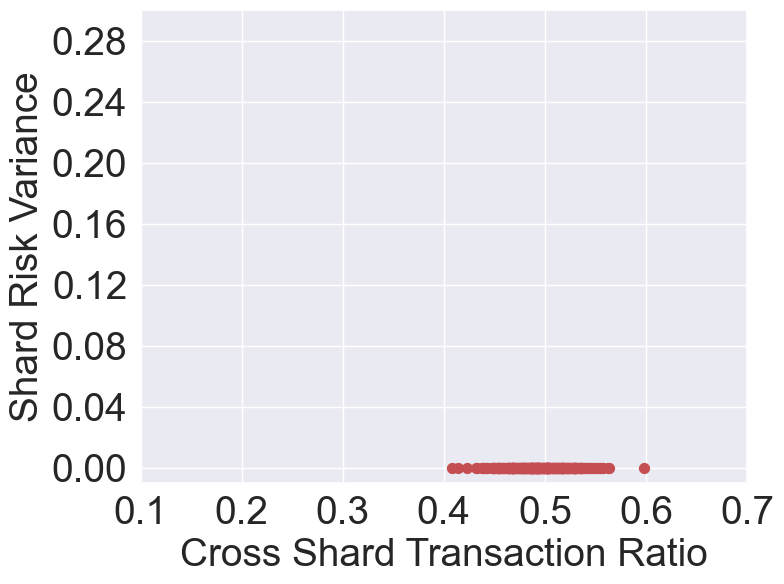}}

    \subfigure[Random-based]{\label{fig.6f}\includegraphics[width=0.19\linewidth]{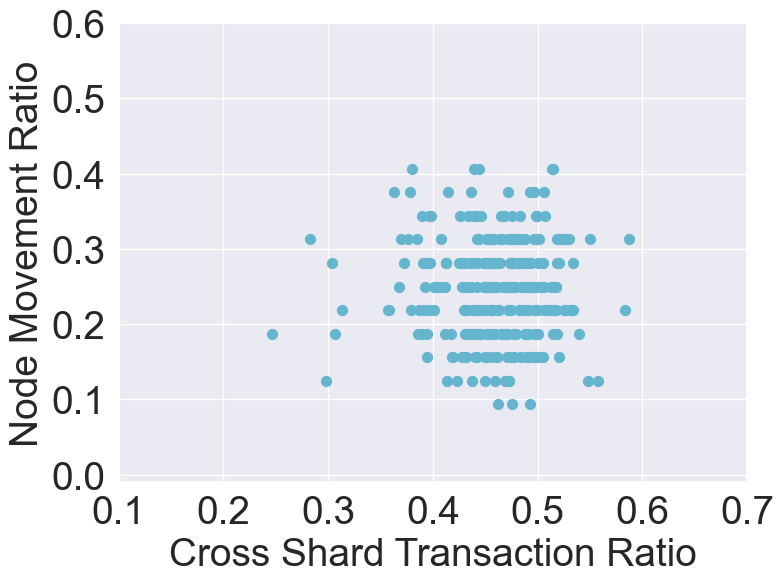}}
    \subfigure[Community-based]{\label{fig.6g}\includegraphics[width=0.19\linewidth]{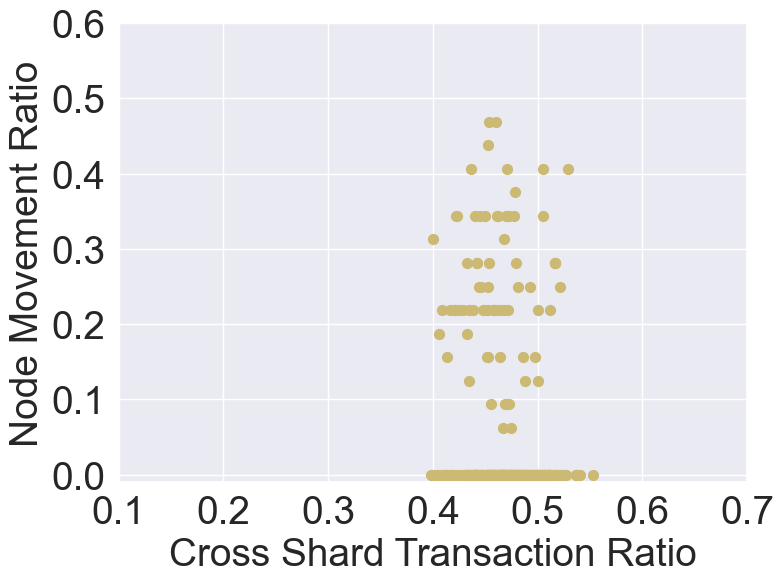}}
    \subfigure[Trust-based]{\label{fig.6h}\includegraphics[width=0.19\linewidth]{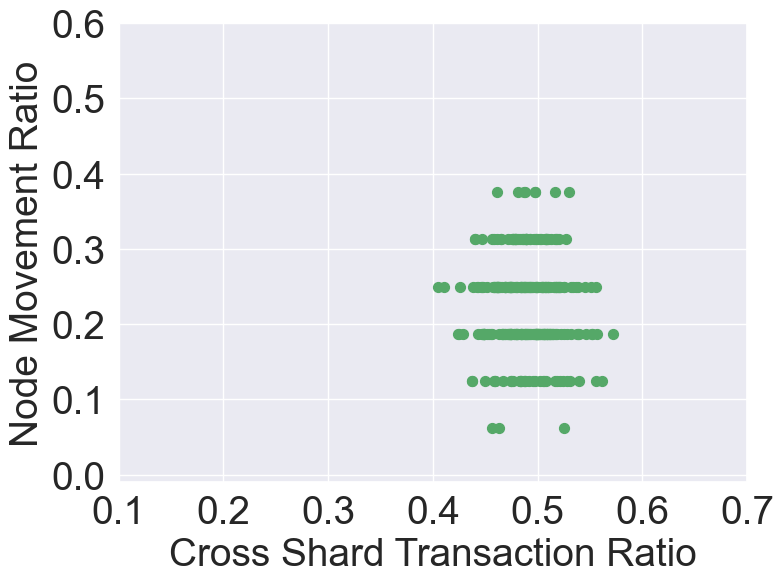}}
    \subfigure[TBDD-DQN]{\label{fig.6i}\includegraphics[width=0.19\linewidth]
    {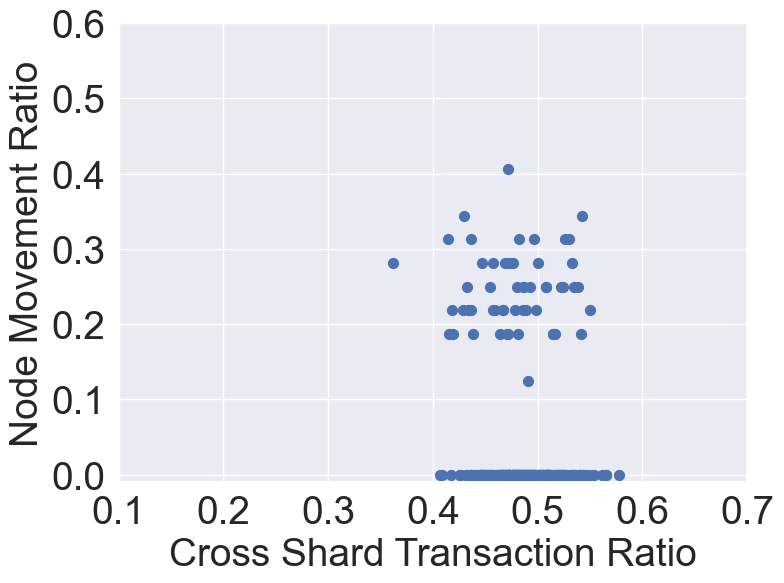}}
    \subfigure[TBDD-PPO]{\label{fig.6j}\includegraphics[width=0.19\linewidth]
    {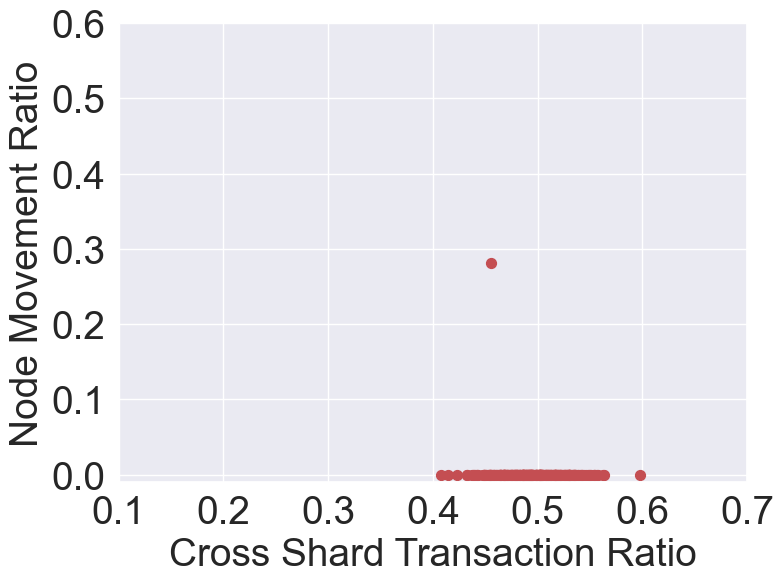}}

    \caption{The comparison of different sharding schemes with same setting $N=16, D=2, h=4$ across $100$ epochs. }
    \label{fig.6}
\end{figure*}

\begin{figure}[h]
    \centering
    \setlength{\belowcaptionskip}{-0.3cm} 
    \subfigure[$N=16, D=2, h=\lbrace0,1,2,3,4,5\rbrace$, TBDD-DQN]{\label{fig.3DQN}\includegraphics[width=0.48\columnwidth]{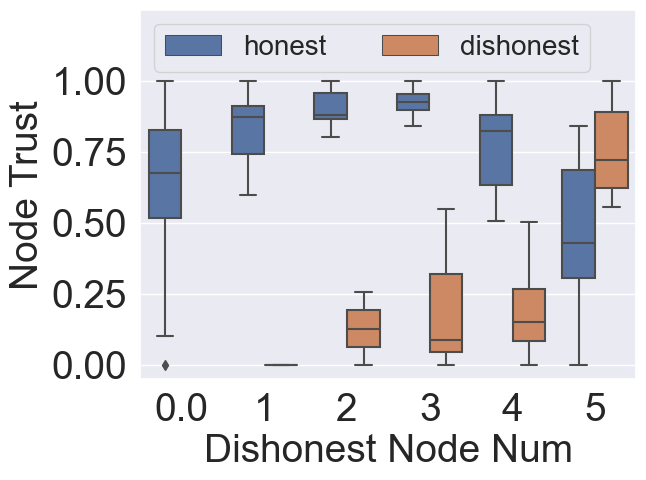}}
    \subfigure[$N=16, D=2, h=\lbrace0,1,2,3,4,5\rbrace$, TBDD-PPO]
    {\label{fig.3PPO}\includegraphics[width=0.48\columnwidth]{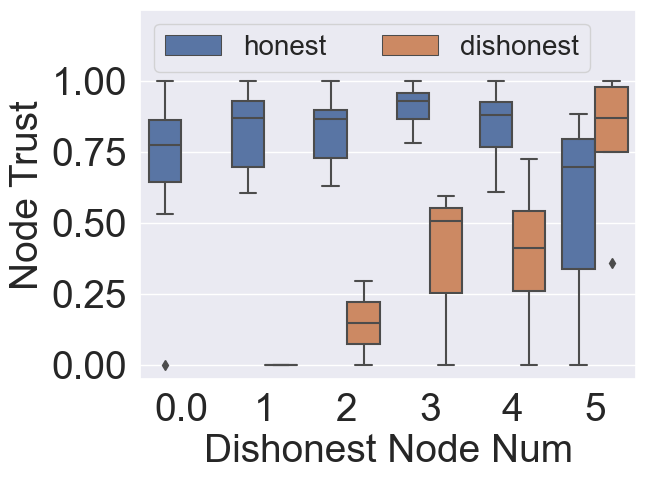}}
	\caption{Tracing of the impact of the number of dishonest nodes for node trust with DRL approach, $N=16, D=2, h=\lbrace0,1,2,3,4,5\rbrace$. The left subfigure uses the DQN algorithm. The right subfigure uses the PPO algorithm.}
	\label{fig.7}
\end{figure}

Fig.~\ref{fig.5a} illustrates the average rewards achieved in a single epoch with different sharding schemes. The community-based sharding technique recorded the lowest reward. While this scheme can significantly minimize cross-shard transactions, it is vulnerable to adaptive collusion attacks. This is because it tends to cluster a high number of dishonest nodes within the same shard, compromising the shard's security. The random-based scheme fares slightly better, with its rewards hovering around zero. The trust-based sharding technique is more proficient at evenly distributing dishonest nodes across different shards, mitigating the risk of a single shard being dominated. However, it leads to a higher number of cross-shard transactions. Consequently, its rewards are marginally less than those of DQN and PPO. Upon reaching the $10^{th}$ epoch, both \textsc{TBDD-PPO} and \textsc{TBDD-DQN} consistently outperform the other sharding schemes, exhibiting consistently high rewards. This indicates that the agent received the maximum reward associated with the action. 

Figs.~\ref{fig.5b}--\ref{fig.5c} depict the convergence of rewards under various node numbers for TBDD-DQN and TBDD-PPO, respectively. The reward becomes more stable as the number of nodes increases. Across Figs.~\ref{fig.5a}--\ref{fig.5c}, a consistent trend emerges caused by the constrained approach in the policy update process. PPO achieves more stable rewards. The instability of DQN is because it combines the direct value function estimation method and $\epsilon$ greedy exploration strategy.

\begin{figure*}[h]
    \centering
    \vspace{-0.3cm}
    \setlength{\belowcaptionskip}{-0.3cm} 
    \subfigure[h=0]{\label{fig.8a}\includegraphics[width=0.32\columnwidth]{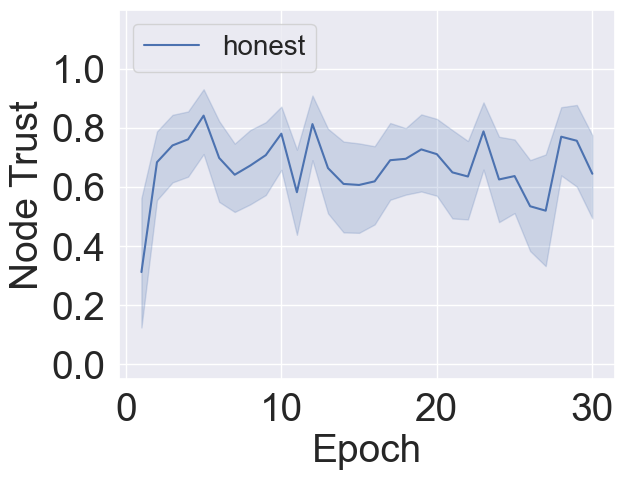}}
    \subfigure[h=1]{\label{fig.8b}\includegraphics[width=0.32\columnwidth]{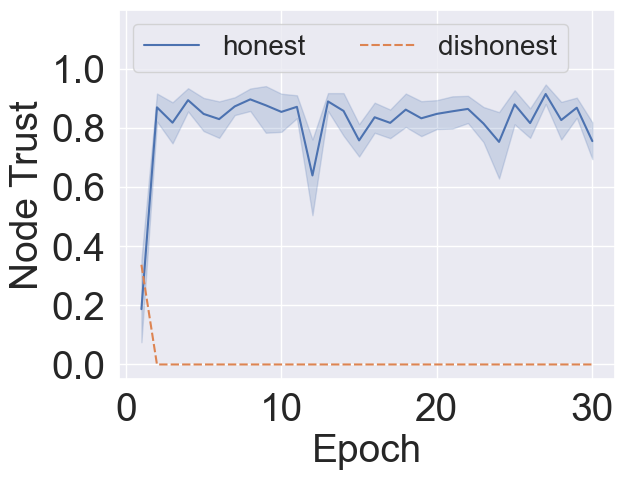}}
    \subfigure[h=2]{\label{fig.8c}\includegraphics[width=0.32\columnwidth]{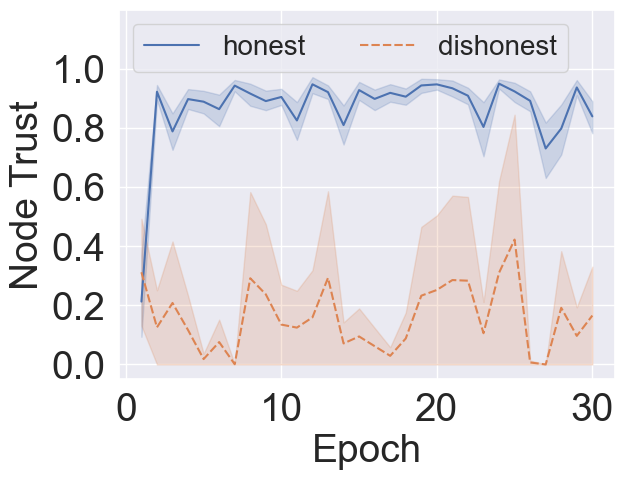}}
    \subfigure[h=3]{\label{fig.8d}\includegraphics[width=0.32\columnwidth]{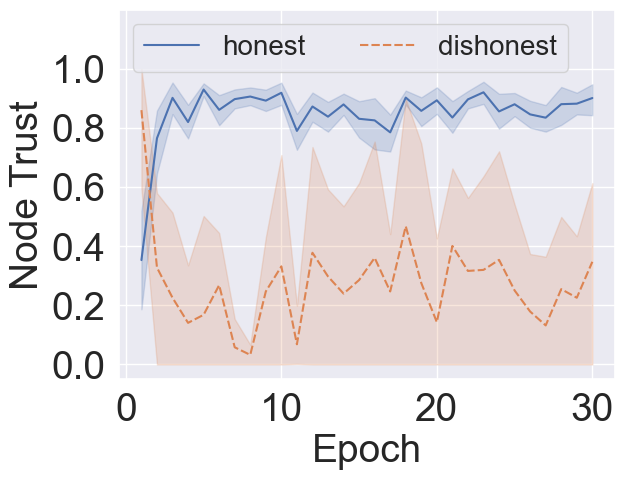}}
    \subfigure[h=4]{\label{fig.8e}\includegraphics[width=0.32\columnwidth]{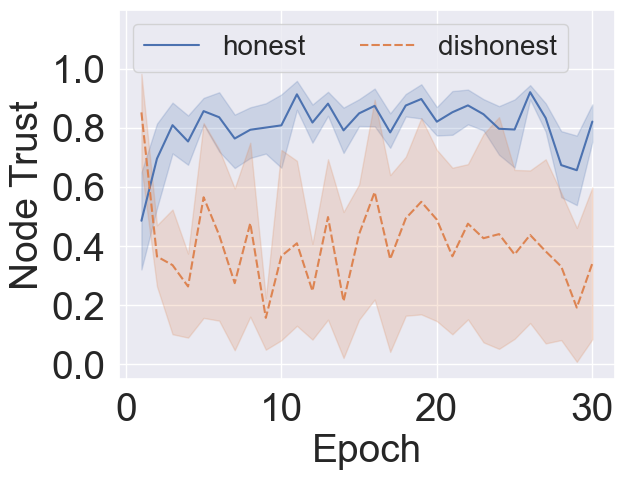}}
    \subfigure[h=5]{\label{fig.8f}\includegraphics[width=0.32\columnwidth]{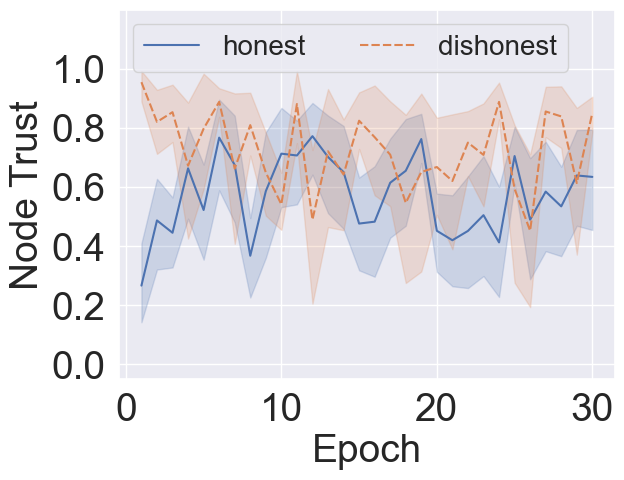}}
    
    \subfigure[h=0]{\label{fig.8g}\includegraphics[width=0.32\columnwidth]{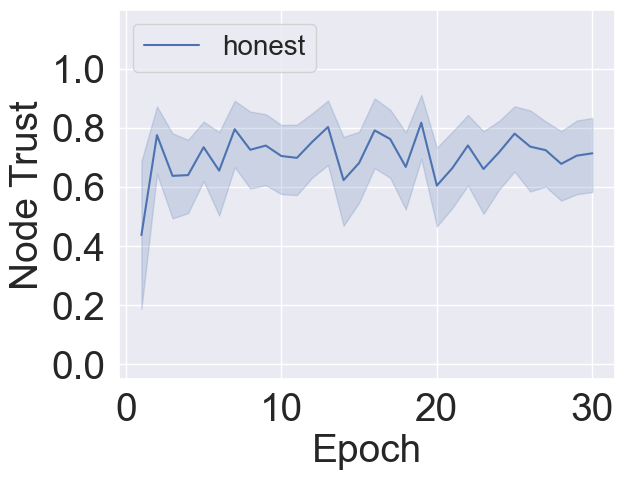}}
    \subfigure[h=1]{\label{fig.8h}\includegraphics[width=0.32\columnwidth]{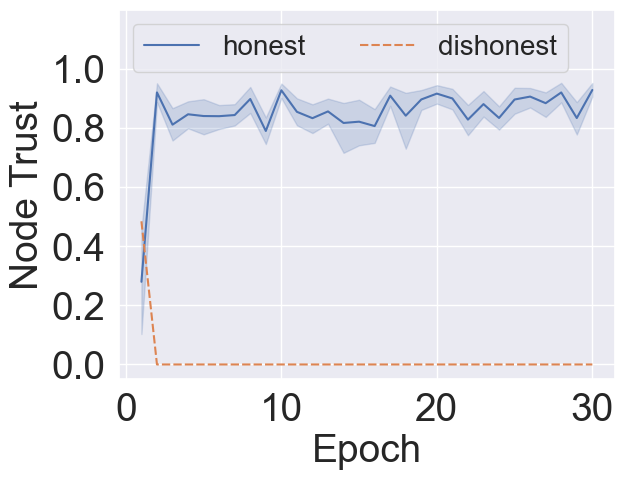}}
    \subfigure[h=2]{\label{fig.8i}\includegraphics[width=0.32\columnwidth]{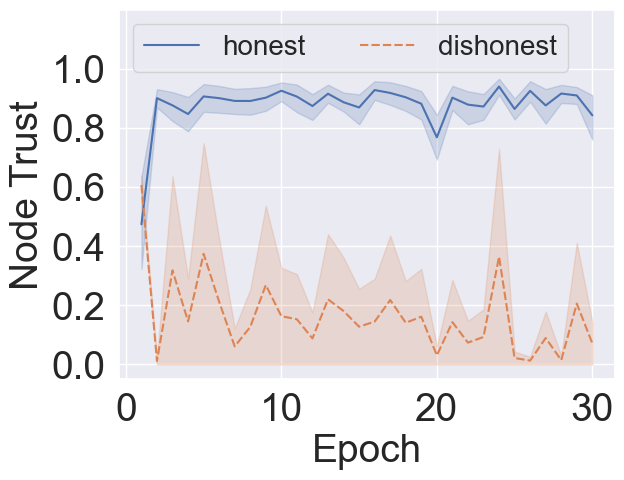}}
    \subfigure[h=3]{\label{fig.8j}\includegraphics[width=0.32\columnwidth]{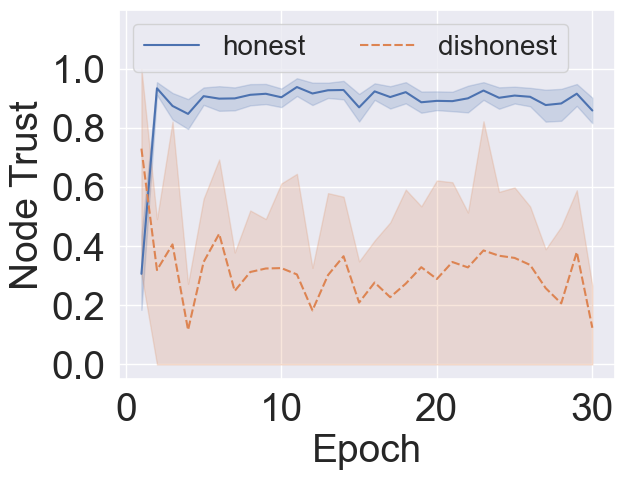}}
    \subfigure[h=4]{\label{fig.8k}\includegraphics[width=0.32\columnwidth]{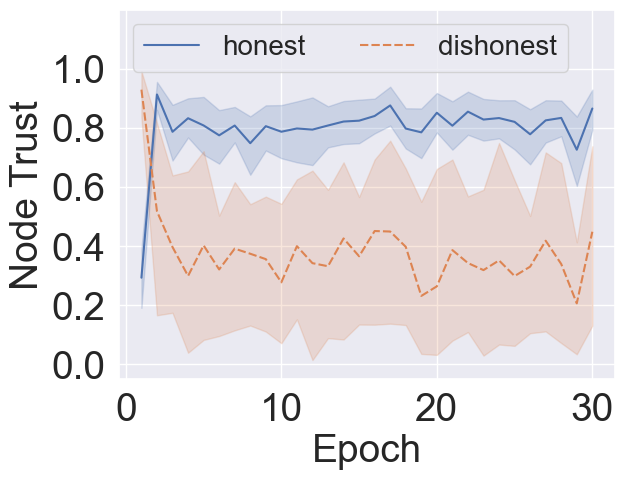}}
    \subfigure[h=5]{\label{fig.8l}\includegraphics[width=0.32\columnwidth]{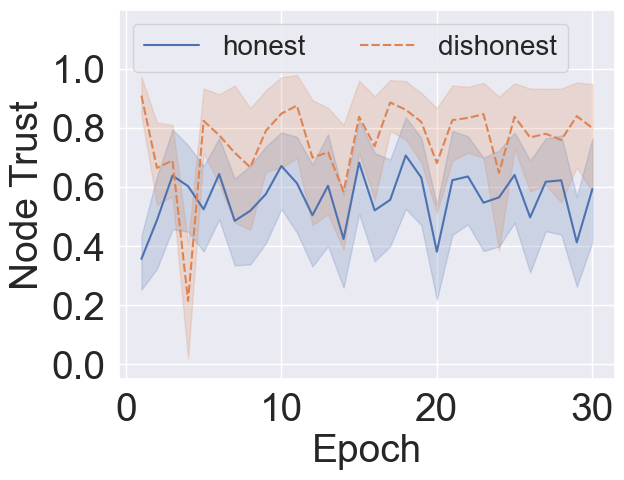}}
    \caption{Tracking global trusts $\mathcal{G}$ between honest and dishonest nodes as the number of dishonest nodes escalates within the DRL scheme. Figs.~\ref{fig.8a}--\ref{fig.8f} represent the trust variation in the DQN algorithm, while Figs.~\ref{fig.8g}--\ref{fig.8l} represent the trust variation in the PPO algorithm. $N=16, D=2, h=\lbrace0,1,2,3,4,5\rbrace$.}
    \label{fig.8}
\end{figure*}

\begin{figure*}[htbp]
    \centering
    \setlength{\belowcaptionskip}{-0.3cm} 
    \subfigure[$N=16, D=2, h=\lbrace0,1,2,3,4,5\rbrace$]{\label{fig.9a}\includegraphics[width=0.44\linewidth]{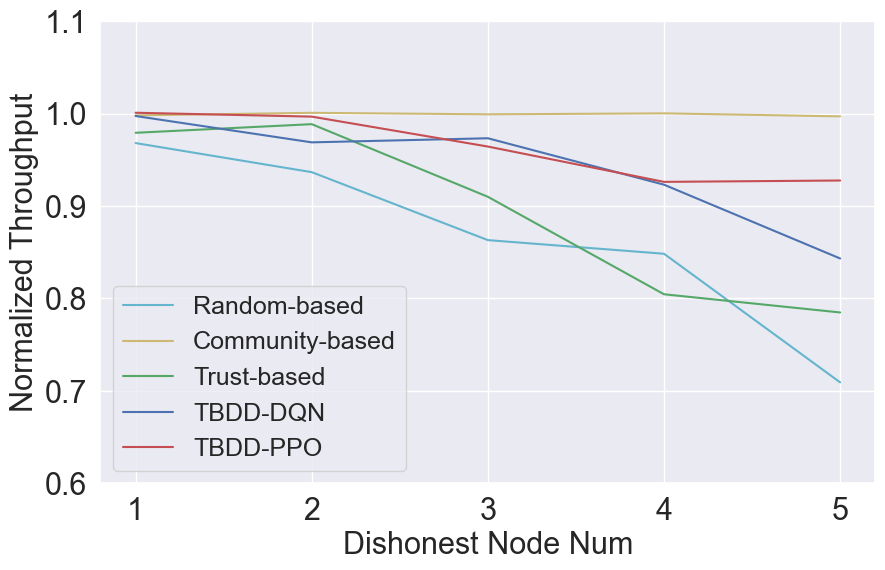}}
    \subfigure[$N=16, D=2, h=\lbrace0,1,2,3,4,5\rbrace$]
    {\label{fig.9b}\includegraphics[width=0.48\linewidth]{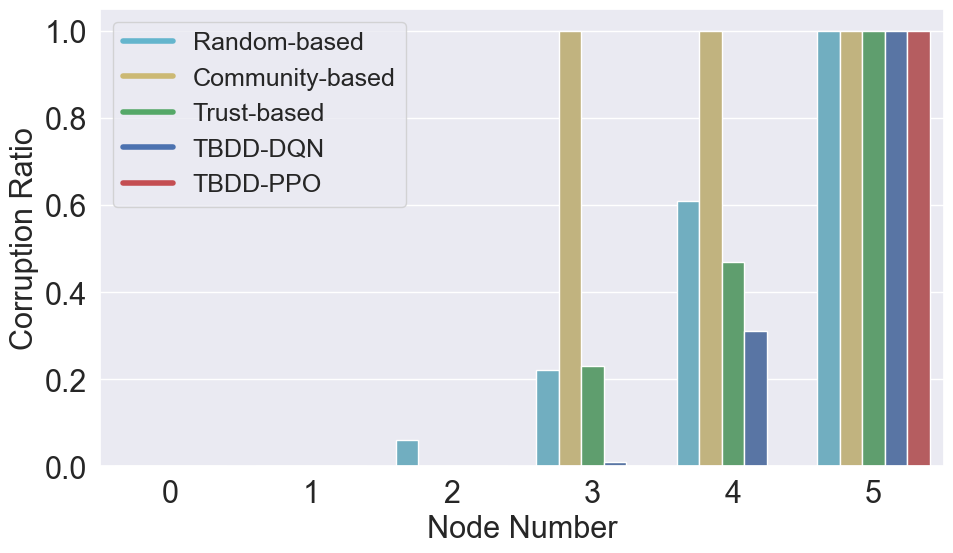}}
    \caption{Fig.~\ref{fig.9a} shows the relationship between the number of dishonest nodes, system throughput, and corrupted shards is interconnected. Fig.~\ref{fig.9b} compares corrupted shard ratio with different sharding techniques across $100$ epochs.}
	\label{fig.9}
\end{figure*}

Figs.~\ref{fig.6a}--\ref{fig.6e} depict the relationship between the CST and shard risk variance. The shard risk variance closer to $0$ indicates a more balanced distribution of malicious nodes, enhancing shard security. The CST closer to $0$ suggests better shard scalability. The results from random-based sharding demonstrate scattered and unpredictable in the scatter plot. Community-based sharding focuses on scalability, overlooking security, leading to a relatively higher shard risk ratio. Trust-based sharding emphasizes security at the expense of scalability, resulting in a relatively larger cross-shard transaction ratio. In contrast, the proposed \textsc{\textsc{TbDd}-DQN} and \textsc{\textsc{TbDd}-PPO} consider both security and scalability. Compared to community-based and trust-based methods, \textsc{\textsc{TbDd}-DQN} and \textsc{\textsc{TbDd}-PPO} are closer to the bottom-left corner, which means that these two sharding methods improve system security while ensuring that the system scalability. This proves the effectiveness of this method, thereby approving the effectiveness of the \textsc{TbDd} framework. Furthermore, consistent with the findings from Fig.~\ref{fig.5}, the \textsc{\textsc{TbDd}-PPO} method demonstrates higher stability than the \textsc{\textsc{TbDd}-DQN} method.

Figs.~\ref{fig.6f}--\ref{fig.6j} illustrate the relationship between CST and node movement ratio. A node movement ratio closer to $0$ indicates fewer node movements, thus reducing system overhead. The outcomes from random-based sharding are unpredictable and uncontrollable. Both Community-based sharding and trust-based sharding methods have a higher number of node movements compared to \textsc{TbDd}-DQN and \textsc{TbDd}-PPO, leading to higher system overhead. Among these, the PPO method results in the fewest node movements. Through Fig.~\ref{fig.6}, we can deduce the following: the \textsc{TbDd}-DQN and \textsc{TbDd}-PPO sharding proposed in this paper achieve an optimal balance between security, scalability, and system overhead, thereby affirming their effectiveness.

In Fig.~\ref{fig.7}, it is shown that the average trust of dishonest nodes surpasses that of honest nodes as the number of dishonest nodes increases. The uppermost horizontal line in each column represents the maximum trust value among all nodes, while the bottom horizontal line represents the minimum trust value. The horizontal line at the midpoint signifies the median trust value of all nodes. When the total number of nodes is $16$, and the total shard number is $2$, the overall fault tolerance threshold $f_{total}$ is $4$ based on \eqref{eq:13}. Consequently, a shard becomes vulnerable and is corrupted when the number of dishonest nodes $h \geq 5$.

As shown in Figs.~\ref{fig.8a}--\ref{fig.8l}, the blue curves represent the average trust of honest nodes, and the red curves are dishonest nodes. Consistent results from Fig.~\ref{fig.7} reveal that the system can effectively perform sharding if the proportion of dishonest nodes remains below $f_{total}$ of the total node count. However, if the number of dishonest nodes exceeds the $f_{total}$ threshold, any sharding approaches, including the proposed schemes \textsc{\textsc{TbDd}-DQN} and \textsc{\textsc{TbDd}-PPO}, cannot safely execute sharding and are vulnerable to the $1\%$ attack, which is beyond the scope of our investigation. On the other hand, having a lower count of dishonest nodes still enables the implementation of more shards within the system, leading to improved overall performance and scalability.

To evaluate throughput across different sharding methods influenced by dishonest nodes, we use the normalized throughput metric. This metric compares the throughput with and without dishonest nodes. A higher ratio suggests dishonest nodes have minimal impact on transaction throughput, while a lower one indicates a significant negative effect. Additionally, we highlight the benefits of our system by comparing the shard corruption ratio over the last $100$ rounds among various sharding approaches. As depicted in Figs.~\ref{fig.9a}--\ref{fig.9b}, random-based sharding exhibits the lowest throughput and compromised security. As dishonest node counts rise, there's a pronounced decline in scalability, coupled with an increased number of corrupted shards. The Community-based sharding method has the best scalability and maintains a high throughput. Yet, its oversight on the security front results in a higher rate of shard corruption. In comparison, the trust-based sharding method reduces the chances of shard corruption but lags in throughput, signaling its scalability constraints. Without collusion attacks and within the tolerable limit of dishonest nodes, the proposed \textsc{\textsc{TbDd}-DQN} and \textsc{\textsc{TbDd}-PPO} in this article achieves approximately a $10\%$ throughput improvement over random-based sharding method and a $13\%$ increase compared to the trust-based method.

\subsection{Discussion}

\noindent\textbf{Latency.}
In \textsc{TbDd}, managing latency is crucial across block verification, sharding strategy optimization, and resharding phases. Tolerating delay during block verification is a necessary trade-off to prevent double-spending issues~\cite{nakamoto2008bitcoin}. However, a prolonged offline DRL learning phase for the sharding strategy is undesirable. 
This could be mitigated by synchronizing online sharding strategy learning with resharding execution. Latency concerns during resharding, due to extensive node synchronization, can be alleviated through state channels~\cite{miller2019sprites} that expedite off-chain transactions, reducing blockchain load and hastening resharding.
Integrating state channels into \textsc{TbDd} enables certain IoT transactions to be executed off-chain during proposed block verification, which lessens the node verification load and boosts overall system responsiveness.

\smallskip
\noindent\textbf{Edge-Driven Protocol.}
The architecture of \textsc{TbDd} uniquely operates on edge servers, not directly on IoT sensor end nodes. This strategic placement ensures that the system can manage extensive computations associated with blockchain operations without overwhelming individual IoT devices. As a result, the scaling exhibited in our experiments is consistent and realistic, representing a practical implementation in real-world IoT networks. This edge-driven approach aligns with the broader move towards edge computing in IoT, capitalizing on its benefits to improve scalability and responsiveness.

\smallskip
\noindent\textbf{Decentralized Coordination.}
Our design leverages a decentralized TC, acting as a trustworthy third-party coordinator, eliminating the vulnerabilities associated with a single centralized coordinator. This approach significantly mitigates the risks of a single point of failure in the system. The intra-consensus security within the decentralized TC ensures robust and reliable decision-making processes. Such a structure has been mirrored in existing studies~\cite{luu2016secure, kokoris2018omniledger}, affirming its practicality and security. By embedding this design, \textsc{TbDd} further ensures system robustness, sustaining its promises of trustworthiness and integrity in diverse IoT settings.

%% file: Sections/6-Conclusions.tex
\section{Conclusion}
\label{section: conclusion}
We introduced \textsc{TbDd}, a novel trust and DRL-based sharding framework for IoT contexts. \textsc{TbDd} surpasses random, community, and trust-based methods by optimally balancing security and scalability, and robustly combating strategic collusion attacks. We devised a trust table mechanism that evaluates collusion, drawing from distributed voting, consensus, and node behavior analysis. Our framework sets dynamic security thresholds, with the decentralized committee acting as an agent that optimizes node allocation by maximizing rewards. Our comprehensive experiments substantiate \textsc{TbDd}'s potential as a robust solution for enhancing security and scalability in real-world IoT environments.